\documentclass[sigconf,natbib=true,authorversion=true]{acmart}

\usepackage[export]{adjustbox}
\usepackage{graphicx}
\usepackage[show]{chato-notes}
\usepackage{epsfig}
\usepackage{algorithm}
\usepackage{algpseudocode}
\usepackage{url}
\usepackage{tikz}
\usepackage[most]{tcolorbox}
\usetikzlibrary{decorations.pathreplacing}
\usetikzlibrary{shadows}
\usepackage{xcolor}
\usepackage[framemethod=TikZ]{mdframed}

\usepackage{type1cm}     %
\usepackage{graphicx}     %
\usepackage{xspace}     %
\usepackage{balance}     %
\usepackage{booktabs}     %
\usepackage{multirow}     %
\usepackage[font={bf}, tableposition=top]{caption}     %
\usepackage{bold-extra}     %
\usepackage{microtype}    %
\usepackage{units}     %
\usepackage{mathtools}     %
\usepackage{subcaption}
\usepackage{enumerate}

\usepackage{dsfont}		%
\usepackage{mathtools,cancel}	%

\newcommand{\spara}[1]{\smallskip\noindent\textbf{#1}}

\newenvironment {squishlist}
{\begin{list}{$\bullet$}
  { \setlength{\itemsep}{0pt}
     \setlength{\parsep}{3pt}
     \setlength{\topsep}{3pt}
     \setlength{\partopsep}{0pt}
     \setlength{\leftmargin}{1.5em}
     \setlength{\labelwidth}{1em}
     \setlength{\labelsep}{0.5em} } }
{\end{list}}

\usetikzlibrary{bayesnet}

\hypersetup{
    bookmarks=true,         %
    unicode=false,          %
    pdftoolbar=true,        %
    pdfmenubar=true,        %
    pdffitwindow=false,     %
    pdfstartview={FitH},    %
    pdftitle={My title},    %
    pdfauthor={Author},     %
    pdfsubject={Subject},   %
    pdfcreator={Creator},   %
    pdfproducer={Producer}, %
    pdfnewwindow=true,      %
    colorlinks=true,       %
    linkcolor=black,          %
    citecolor=black,        %
    filecolor=black,      %
    urlcolor=black           %
}

\DeclareMathOperator*{\argmax}{arg\,max}

\renewcommand{\algorithmicrequire}{\textbf{Input:}}
\renewcommand{\algorithmicensure}{\textbf{Output:}}

\newcommand{\signfun}{\mathcal{S}}
\newcommand{\sigmoid}{f}

\newcommand{\indep}{\mathrel{\text{\scalebox{1.07}{$\perp\mkern-10mu\perp$}}}}
\newcommand{\nindep}{\cancel{\mathrel{\text{\scalebox{1.07}{$\perp\mkern-10mu\perp$}}}}}

\newcommand{\mup}{\ensuremath{\mu^{+}}\xspace}
\newcommand{\mun}{\ensuremath{\mu^{-}}\xspace}
\newcommand{\epsp}{\ensuremath{\epsilon^{+}}\xspace}
\newcommand{\epsn}{\ensuremath{\epsilon^{-}}\xspace}

\newcommand{\name}{Learnable Opinion Dynamics Model\xspace}
\newcommand{\nameacr}{LODM\xspace}

\copyrightyear{2020}  %
\acmYear{2020}  %
\setcopyright{acmlicensed} %
\acmConference[KDD '20] {26th ACM SIGKDD Conference on Knowledge Discovery and Data Mining}{August 23--27, 2020}{Virtual Event, USA} %
\acmBooktitle{26th ACM SIGKDD Conference on Knowledge Discovery and Data Mining (KDD '20), August 23--27, 2020, Virtual Event, USA} %
\acmPrice{15.00} %
\acmDOI{10.1145/3394486.3403119} %
\acmISBN{978-1-4503-7998-4/20/08}  %

\ccsdesc[500]{Computing methodologies~Learning in probabilistic graphical models}
\ccsdesc[500]{Computing methodologies~Agent / discrete models}
\ccsdesc[500]{Human-centered computing~Social network analysis}

\graphicspath{{figures/}}

\begin{document}
\fancyhead{} %

\title{Learning Opinion Dynamics From Social Traces}

\author{Corrado Monti}
\affiliation{\institution{ISI Foundation, Italy}}
\email{corrado.monti@isi.it}

\author{Gianmarco De~Francisci~Morales}
\affiliation{\institution{ISI Foundation, Italy}}
\email{gdfm@acm.org}

\author{Francesco Bonchi}
\affiliation{\institution{ISI Foundation, Italy}}
\affiliation{\institution{Eurecat, Spain}}
\email{francesco.bonchi@isi.it}

\begin{abstract}
Opinion dynamics --the research field dealing with how people's opinions form and evolve in a social context-- traditionally uses agent-based models to validate the implications of sociological theories.
These models encode the causal mechanism that drives the opinion formation process, and have the advantage of being easy to interpret.
However, as they do not exploit the availability of data, their predictive power is limited.
Moreover, parameter calibration and model selection are manual and difficult tasks.

In this work we propose an inference mechanism for fitting a generative, agent-like model of opinion dynamics to real-world social traces.
Given a set of observables (e.g., actions and interactions between agents),
our model can recover the most-likely latent opinion trajectories that are compatible with the assumptions about the process dynamics.
This type of model retains the benefits of agent-based ones (i.e., causal interpretation), while adding the ability to perform model selection and hypothesis testing on real data.

We showcase our proposal by translating a classical agent-based model of opinion dynamics into its generative counterpart.
We then design an inference algorithm based on online expectation maximization to learn the latent parameters of the model.
Such algorithm can recover the latent opinion trajectories %
from traces generated by the classical agent-based model.
In addition, it can identify the most likely set of macro parameters used to generate a data trace, thus allowing testing of sociological hypotheses.
Finally, we apply our model to real-world data from Reddit to explore the long-standing question about the impact of the \emph{backfire effect}.
Our results suggest a low prominence of the effect in Reddit's political conversation.

\end{abstract}
\maketitle

\section{Introduction}
\label{sec:intro}

\emph{Opinion dynamics} is the study of how people's opinion on a subject matter form and evolve with time~\citep{french1956formal,harary1959criterion}.
This branch of social psychology has recently received growing attention due to the widespread adoption of social-media platforms. 
Users of these platforms can easily access and consume an immense amount of content, 
as well as engage in debate. In doing so, users share publicly their comments and beliefs, what they like and what they do not like, in other words,
i.e., they leave \emph{data traces}.
Modeling opinion dynamics from this wealth of data is thus a tremendous opportunity for the social scientist. However, traditional opinion dynamics model are \emph{agent-based}, i.e., they are simulations where a set of agents, interconnected by a network, interacts according to pre-determined mechanisms.
These interactions modify the internal opinions of the agents, which in turn generate the dynamic of the opinion formation process.

Starting with the classical model by~\citet{degroot1974reaching}, a plethora of refinements have been proposed~\citep{friedkin1990social,axelrod1997dissemination,deffuant2000mixing}, all
sharing the fundamental strengths and weaknesses of agent-based models (ABMs)~\citep{squazzoni2012agentbased}. ABMs offer a framework for theory development, by allowing to explore empirically the implications of a sociological hypothesis formalized as a rule for interaction among agents. 
As such, ABMs provide a mechanistic model, which is easily interpretable in a causal way.
This property is in sharp contrast with other models used in social science, such as statistical models (e.g., regression), for which a causal interpretation is much harder~\citep{pearl2009causal}.
However, agent-based models also have several shortcomings.
First, their predictive power is rather limited~\citep{deffuant2008agent}.
Second, parameter calibration is a considerable challenge, as it needs to be performed largely by hand.
Third, agents cannot be directly used to understand any individual-level digital trace (e.g., from the Web or social media).
Typically, in fact, \emph{ABMs do not involve any inference from data}.

In this paper, we overcome these shortcomings of ABMs by proposing \emph{an inference mechanism for fitting a generative, agent-like model of opinion dynamics to real-world social traces}.
Such a model, dubbed \emph{\name} (\nameacr), retains the desirable properties of ABMs (causal interpretation of the mechanism behind opinion dynamics), while at the same time allowing for parameter inference from real data.
Consequently, it can be used to explain individual behaviors, for model selection, and even for prediction: in other words, it produces a more \emph{testable hypothesis}.

In particular, we translate a classical agent-based opinion dynamics model by~\citet{jager2005uniformity} into a probabilistic generative framework.
This classical model relies on \emph{bounded confidence} with a \emph{backfire} extension, based on social judgment theory~\citep{sherif1961social}.
After translating the model, we design an inference algorithm, based on online expectation maximization, that can fit the model \emph{micro-level} parameters, the opinions of the agents, by looking at a data trace.
We show how to use our framework for model selection, i.e., to identify the most likely \emph{macro level} parameters of the model, the rules which prescribe the opinion dynamics, from the data.

Finally, we apply our proposed model to a real-world dataset from Reddit, containing 
10-years longitudinal cross-section of active users on subreddits related to 
politics. We show that our model is able to capture several behaviors of online users,
such as the popularity of a user within a community, and the emergence of conflicts between users.
Moreover, we show how our framework can test concrete sociological hypotheses expressed as agent-based interaction rules.
In particular, we use the model to answer the question \emph{``is there evidence of backfire effect in political discussion on Reddit?''},
to which we find a negative answer.

\section{Related work}
\label{sec:related}

Opinion dynamics models (ODM) deal with the evolution over time of opinions in groups~\citep{coates2018unified}, and study sociological phenomena such as consensus formation~\citep{degroot1974reaching}, attitude change~\citep{sherif1961social, jager2005uniformity} and polarization~\citep{delvicario2017modeling}.
One of the most popular~\citep{mathias2016bounded, gomez-serrano2010bounded} continuous-valued models is the bounded confidence model (BCM) by~\citet{deffuant2000mixing}, which explains the observed differences in opinion through \emph{bounded confidence}:
agents ignore what is perceived as too distant from their own beliefs.
 Several extensions of BCM have been proposed by implementing other observations from sociology~\citep{castellano2007statistical}.
\citet{quattrociocchi2011opinions} employed social impact theory, which emphasize the role of group pressure in attitude change.
\citet{jager2005uniformity} instead built on social judgment theory~\citep{sherif1961social}: the result of persuasion depends critically on the position of the receiver, and could end up with \emph{acceptance} or \emph{contrast} (the latter also known as \emph{backfire effect}).
The backfire effect suggests a link between exposition to opposing views (e.g., on social media) and polarization; as such, its importance has recently become a widely debated issue.
Both \citet{sippitt2019backfire} and \citet{fletcher2019polarisation} noted the need for more empirical tests confirming or disproving the backfire effect.

In fact, a great concern in opinion dynamics is how to validate the results empirically.
According to~\citet{flache2017modelsa}, the field suffers from ``a proliferation of theoretical studies and a dearth of empirical work''; for~\citet{castellano2007statistical}, ``there is a striking imbalance between empirical evidence and theoretical modelization, in favor of the latter''.
Therefore, the question of empirical validation has recently started to attract some attention.
For instance, \citet{sobkowicz2016quantitative} try to calibrate the model in order to reproduce some observations on the distribution of the resulting opinions in the population.
This method has been described~\citep{flache2017modelsa} as a test on \emph{macro-level} predictions.
Instead, the connection of \emph{micro-level} (i.e., individual agents) behavior with real-world observations, that we tackle in this paper, is still largely unexplored.

Some of our ideas are also found in recent work, albeit within different conceptual frameworks, and with different techniques.
The inclusion of actions as an observable for opinions was proposed also by~\citet{tang2019learning}, but without taking into account statistical inference nor real-world observations.
Estimation through Maximum-a-Posteriori was proposed by~\citet{sichani2017inference} for the sole purpose of inferring the most influential nodes; they considered the opinions to be fully observable.

\citet{de2016learning} used bayesian inference coupled with an ad-hoc model to predict opinion diffusion trough social influence.
They assume that the polarity of messages is given and that followers of a user can be influenced by her messages.
In particular, each observable is a triplet $(u,m,t)$, indicating that the user $u$ posted a message with sentiment $m$ at time $t$, while in our model we observe interactions among users (e.g., discussion) and actions performed by users (e.g., sending a message) without a predefined polarity for the actions.
Finally,~\citet{grazzini2017bayesian} proposed bayesian estimates to calibrate the parameters of other ABMs (not opinion dynamics), but do not consider micro-level predictions.

\section{Generative framework}
\label{sec:framework}

In agent-based opinion dynamics models, interactions between agents are the driver of opinion change.
For instance, the model by~\citet{jager2005uniformity} distinguishes different kinds of interactions (positive and negative) with opposite effects on the opinions of the involved agents.
In reality, neither the opinion of a single agent nor the ``sign" of the interactions are easily observable.
Therefore, it is difficult to use such models to explain individual behavior.

\begin{figure}[t]
\begin{center}
\includegraphics[width=\columnwidth]{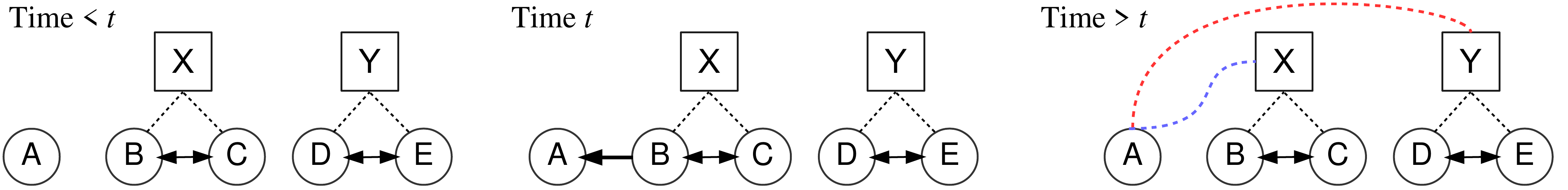}
\vspace{-2mm}
\caption{A minimal example of the observed actors (circles) and actions (squares), in the case of a positive (blue) and negative (red) interaction.
}
\label{fig:example}
\end{center}
\vspace{-2mm}
\end{figure}

What is observable, instead, is that an interaction between two agents has happened.
Moreover, we can often observe some \emph{action} performed by individuals: using a hashtag on Twitter, or participating in a specific Reddit community. %
Such actions are often a reflection of an individual's personal opinion: hashtags are used as propaganda tools by political campaigns (e.g., \texttt{\#MAGA}), Reddit communities gather people with similar views (e.g., \texttt{r/The\_Donald}).

Our proposed probabilistic framework \nameacr aims exactly at explaining the individual behaviors recoverable from the digital traces found in social media.
Our goal is therefore to estimate the micro-level latent variables of interest (i.e., individual opinions, the sign of interactions) given the observed ones (i.e., interactions and actions), under the assumptions of a specific opinion dynamics model.
It is thus natural to frame our problem as a probabilistic generative model: the input are observed variables, the output are estimates for the latent ones.

\subsection{Observables}
\label{sec:observables}
Let $V$ be a set of \emph{actors}, who interact and influence each other's opinion.
We represent interactions as an arc in a temporal graph, defined over $T$ discrete time steps, where each actor is a node.
Actors also perform \emph{actions}.
Each action is driven by the latent opinion of the actor: different opinions lead to different actions (think, for instance, of putting a ``like'' on a politically-charged Facebook page).
We consider actions as a noisy proxy for the opinion of an actor.
Let $A$ be the set of possible actions.
We represent the fact that an actor performs an action as a temporal arc in a bipartite graph, defined by $V$ and $A$.

Formally, we observe the following two temporal graphs:
\begin{squishlist}

  \item $G=(V, E)$ is the directed interaction graph between actors.
  Arc $(u, v,  t) \in E$ represent that ``$u$ interacts with $v$ at time $t$''.
  The interaction results in $u$ possibly influencing $v$.
  Actors can interact multiple times at time step $t$, so we define $E$ as a multiset, and $G$ as a multigraph.

  \item $Z=(V, A, F)$ is a bipartite graph of actors and actions.
  Arc $(v, a, t) \in F$ represents that ``actor $v$ performs action $a$ at time $t$''. Similarly to $G$, each arc can appear multiple times in the same time step.
  Therefore, $Z$ is a bipartite multigraph.

\end{squishlist}

We depict a minimal example of these observables in Figure~\ref{fig:example}.
We have $5$ actors (A,B,C,D,E), and $2$ possible actions (X,Y);
Actors B and C perform action X at all time steps, while actors D and E perform action Y.
These conditions create two clusters of actors who do not interact with each other: a typical instance of polarized opinions.
Actor A plays the central role, as its action changes after interacting with the other actors.
In the \emph{consensus} scenario, A interacts with B at time $t$, and then performs action $X$ from time $t+1$ onwards.
In the \emph{backfire} scenario, A interacts with B at time $t$, but then performs action $Y$ from time $t+1$ onwards.

\subsection{Latent variables}
In our setting, each interaction in $G$ is either positive or negative, and it changes the latent opinions of the actors accordingly.
The idea that interactions can have different effects is a key concept in several opinion dynamics models \citep{delvicario2017modeling, allahverdyan2014opinion, jager2005uniformity, chen2019opinion, stefanelli2014moderate}.
In addition, we need to represent actions in opinion space.
Each action is associated to a range of opinions, fixed in time for simplicity.
We express these concepts via the following latent variables:
\begin{squishlist}
  \item $x_{t, v} \in [-1, 1]$ represents the latent opinion of actor $v \in V$, on a given subject
matter, at time $t$.
  \item $\signfun: E \rightarrow \{-1, +1\}$ represents the signs of the interaction arcs, which characterize each interaction between actors as either positive or negative.
  \item $w_{a} \in [-1, 1]$ and $\sigma_a$ are the center and half-width of the opinion spectrum $[w_a - \sigma_a, w_a + \sigma_a]$ associated to
action $a \in A$.
\end{squishlist}

\subsection{Base model}
\label{sec:model}
Next, we describe the original, deterministic ABM by~\citet{jager2005uniformity}.
This model assumes that interactions are either positive or negative.
This is determined by two macro parameters:
a \emph{latitude of acceptance} and a \emph{latitude of contrast}, denoted with \epsp and \epsn respectively\footnote{The original
paper uses the notation $T$ and $U$.} (s.t. $0 \leq \epsp < \epsn \leq 2$).
The sign of an interaction $(u,v,t) \in E$ is determined when $u$ expresses its opinion to $v$: if it is close (within \epsp) $v$ \emph{accepts} it, if it is distant (further than \epsn) $v$ \emph{constrasts} it. $\forall (u,v,t) \in E$
\begin{multline}
  \begin{split}
    |x_{t, u} - x_{t, v}| < \epsilon^+
      & \implies
      & \signfun(u, v, t) = +1
      & \text{ (positive arc) }
\\
    |x_{t, u} - x_{t, v}| > \epsilon^-
      & \implies
      & \signfun(u, v, t) = -1
      & \text{ (negative arc), }
\\
  \end{split}
  \label{eq:deterministic-arcs}
\end{multline}
and the opinions are updated accordingly
\begin{multline}
  \begin{split}
    \signfun(u, v, t) = +1
      & \implies
      & x_{t + 1, v} = x_{t, v} + \mu^+ \cdot (x_{t, u} - x_{t, v}) \\
    \signfun(u, v, t) = -1
      & \implies
      & x_{t + 1, v} = x_{t, v} - \mu^- \cdot (x_{t, u} - x_{t, v}), \\
  \end{split}
  \label{eq:opinion-update}
\end{multline}
while clipping $x_{t + 1, v}$ in $[-1, 1]$. The parameters $\mup, \mun > 0$ thus control the speed of the influence due to the interactions.

\subsection{Generative process for interactions}
\label{sec:generate-interactions}

Next, we describe how we translate this deterministic ABM into its probabilistic generative counterpart.
This change allows us to design an inference procedure for the latent variables, via maximum a posteriori likelihood estimation.
The modified model maintains the deterministic update rules for opinions of the agents~(Eq.~\ref{eq:opinion-update}).

To make the generative model realistic, there are a few technical concerns to address.
An actor might have more interactions in some time steps and fewer in others, for exogenous reasons.
In addition, some actors might, in general, interact more than others.
We wish for our model to keep these concerns into account, but without modeling them explicitly.
Therefore, in our model (1)~a node $u$ at time $t$ generates a given, fixed number $\gamma_{t, u}$ of arcs;
(2)~at each time step $t$, only a subset $V^*_t \subset V$ of the nodes is considered active and eligible to receive an arc.
In other words, we do not explicitly model the probability of directly drawing an arc from all possible pairs given the opinions of agents $P\big( (u, v, t) \in E \mid \mathbf{x}_{t} \big)$.
This design choice allows the model to accept any real interaction graphs, with any observed empirical degree distribution, %
similarly to the configuration model~\citep{bender1978asymptotic}.

In order to make interactions stochastic, we first need to determine the a priori probability of an interaction being positive at time $t$. Considering the opinions $\mathbf{x}_{t}$, we can use a summary statistic: the fraction of possible positive interactions

\begin{small}
\begin{equation}
  \label{eq:alphat}
\alpha_t = \frac{\sum\limits_{(u,v) \in E_t} { \mathds{1} \left( |x_{t,u} - x_{t,v}| < \epsp \right) } }
              {
               \sum\limits_{(u,v) \in E_t} { \mathds{1} \left( |x_{t,u} - x_{t,v}| < \epsp \right) } +
               { \mathds{1} \left( |x_{t,u} - x_{t,v}| > \epsn \right) } .
              }
\end{equation}
\end{small}
Given $u$, to draw one of the $\gamma(u,t)$ arcs, we first draw a sign for the arc, positive with probability $\alpha_t$, and then we pick the target $v$ among the available nodes within the latitude for the given sign.

Now, to turn the agent-based model into a probabilistic generative one, we wish to add stochastic behavior into Equation~\ref{eq:deterministic-arcs}.
In particular, to account for noise in the data, we relax the boundaries on the latitudes.
We define the probability of an interaction $(u, v, t)$ as a function of the opinions of $u$ and $v$, and of the sign of the arc.
Let $\sigmoid_G(x) = 1 / (1 + e^{-\rho_G \cdot x})$ be a sigmoid function with a certain steepness $\rho_G$.
Then, we define the probability of an interaction so that it depends on two functions
\begin{multline}
  \begin{split}
    \kappa^+(x_{t, u}, x_{t, v})
      & := \sigmoid_G \left( \epsilon^+ - |x_{t, u} - x_{t, v}| \right)
      & \text{(positive)}\, \\
    \kappa^-(x_{t, u}, x_{t, v})
      & := \sigmoid_G \left( |x_{t, u} - x_{t, v}| - \epsilon^- \right)
      & \text{(negative)}. \\
  \end{split}
  \label{eq:probability-arcs}
\end{multline}

We can now use these functions to define a probabilistic generative process for the observed temporal graph, such that
\begin{multline}
  \begin{split}
    P \big( (u, v, t) \in E \mid \signfun(u, v, t) = +1 \big) & \propto \kappa^+(x_{t, u}, x_{t, v}) \\
    P \big( (u, v, t) \in E \mid \signfun(u, v, t) = -1 \big) & \propto \kappa^-(x_{t, u}, x_{t, v}).
  \end{split}
  \label{eq:probability-arcs-propto}
\end{multline}
As the steepness of the sigmoid goes to infinity, Equations~\ref{eq:probability-arcs}~and~\ref{eq:probability-arcs-propto} turn into the original opinion dynamics model (Equation~\ref{eq:deterministic-arcs})~\cite{jager2005uniformity}, where every node within the latitude is equally likely to interact with the originating node, and all the nodes outside the latitude have zero probability of interacting with it.

In summary, we define the following overall generative process. For each time step $t$:
\begin{enumerate}[(i)]
  \item Determine $\alpha_t$ from $\mathbf{x}_t$.
  \item For each actor $u$, for $\gamma_{t, u}$ times:
  {\setlength\itemindent{25pt} \item Extract a sign $s \in \{ -1, +1 \}$ with probability $\alpha_t$.}
  {\setlength\itemindent{25pt} \item Choose an actor $v \in V^*_t$ with probability:}
  \begin{equation}
    \label{eq:link-probability}
    P((u, v, t) \in E \mid u, \mathbf{x}_t, s) = \frac{\kappa^s(x_{t, u}, x_{t, v})}
                     {\sum_{v' \in V^*_t} \kappa^s(x_{t, u}, x_{t, v'}) }
  \end{equation}
  {\setlength\itemindent{25pt} \item Add the interaction $(u, v, t)$ to $E$.}
  \item Finally, update $x_{t + 1, v}$ according to Equation~\ref{eq:opinion-update}.
\end{enumerate}

\subsection{Generative process for actions}
\label{sec:generative-features}

We now define a similar process to account for the \emph{actions} performed by each actor.
Let $\zeta_{t, u}$ to be the exogenous, given number of actions that node $u$ performs at time $t$.
We define $P \big( (u, a, t) \in F \big)$ to be proportional to
\begin{equation}
  \kappa_\sigma(x_{t, u}, w_a) :=
    \sigmoid_Z \big( \sigma_a - |x_{t, u} - w_a| \big),
\end{equation}
where $\sigmoid_Z$ is a sigmoid function with steepness $\rho_Z$,
and $\sigma_a$ represents a latent concentration in opinion space for action $a$.

Then, we assume that actions are performed at time step $t$ according to the following process:

\begin{enumerate}[(i)]
  \item For each actor $u$, for $\zeta_{t, u}$ times:
{\setlength\itemindent{25pt} \item Choose an action $a$ with probability:}
\begin{equation}
  P((u, a, t) \in F \mid v, \mathbf{x}_t, \mathbf{w}, \boldsymbol{\sigma}) =
              \frac{ \kappa_\sigma (x_{t, u}, w_a)}
                   {\sum_{a' \in A} \kappa_\sigma (x_{t, u}, w_a) }
  \label{eq:feature-probability}
\end{equation}
{\setlength\itemindent{25pt} \item Add performed action $(v, a, t)$ to $F$.}
\end{enumerate}

The described model for actions and interactions is represented via plate notation in Figure~\ref{fig:graphical-model}.
We provide in Appendix~\ref{sec:notation} a reference table outlining the notation we used.

\begin{figure}
  \vspace{-2mm}
\begin{center}
\resizebox{0.52\columnwidth}{!}{%
\begin{tikzpicture}
  \node[det]                          (x)     {$\mathbf{x}_t$};
  \node[latent, above=0.4cm of x]     (x0)    {$\mathbf{x}_0$};
  \node[det, below=2.2cm of x]        (xt1)   {$\mathbf{x}_{t+1}$};
  \node[det, left=2.3cm of x]         (alpha) {$\alpha_t$};
  \node[latent, below=0.6cm of alpha] (s)     {$s$};
  \node[obs, right=0.8cm of s]        (e)     {$u, v$};
  \node[obs, right=2.25cm of e]        (f)     {$v, a$};
  \node[latent, above=1.7cm of f]     (w)     {$\mathbf{w}$};
  \node[latent, right=0.8cm of w]     (sigma) {$\mathbf{\sigma}$};

  \plate [minimum size=3cm, minimum width=6cm] {} {(alpha) (s) (e) (f) (x) (xt1)} {$T$} ;
  \plate [minimum size=1.4cm] {} {(s) (e)} {$\gamma_t$} ;
  \plate [minimum size=1.4cm] {} {(f)} {$\zeta_t$} ;

  \edge {x} {alpha}
  \edge {alpha} {s}
  \edge {s} {e}
  \edge {x} {e}
  \edge {x0} {x}
  \edge {x} {xt1}
  \edge {e} {xt1}
  \edge {s} {xt1}

  \edge {x} {f}
  \edge {sigma} {f}
  \edge {w} {f}
\end{tikzpicture}
}%
\end{center}
  \vspace{-0.5\baselineskip}
  \caption{Plate diagram of our model.%
  }
  \label{fig:graphical-model}
  \vspace{-\baselineskip}
\end{figure}
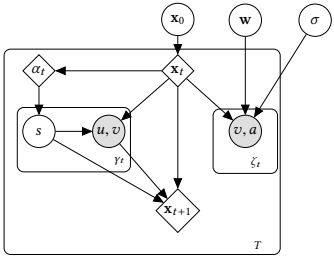

\spara{Example.} In the minimal example discussed at the end of Section~\ref{sec:observables} and depicted in Figure~\ref{fig:example}, this model assumes that the observed behavior is a result of a positive or a negative interaction.
In the positive example, the opinion of the two nodes $A$ and $B$ are likely within the latitude of acceptance (i.e., $|x_A - x_B| < \epsp$).
The interaction between them is therefore positive and brings them closer together by a factor \mup.
Thus, the actor-action arc $(A, X, t)$ that we observe becomes more likely, since $x_A \sim w_X \sim x_B$.

In the negative example, the opinion of the two nodes $A$ and $B$ are likely to be separated at least by the latitude of contrast (i.e., $|x_A - x_B| > \epsn$).
The interaction between them is negative and pushes them apart by a factor \mun.
Thus, the observed actor-actor arc $(A, Y, t)$ is more likely, since the action $Y$ is probably far from $A$ and $B$, who never performed it.

\section{Learning}
\label{sec:algorithms}

Next, we present an algorithm to maximize the complete-data likelihood of the model,
which estimates the latent variables given the observables and the macro parameters.

\subsection{Complete-data likelihood}
We can write the complete likelihood of a given dataset (under the knowledge of all the latent variables) as $P(E, F) = P(E) \cdot P(F)$, thus factoring the likelihood into the interaction likelihood $P(E)$ and the action likelihood $P(F)$.
Note that the process described in Section~\ref{sec:generate-interactions} implies that, by total probability, the interaction likelihood $P(E)$ can be decomposed into the two mutually exclusive cases of positive and negative interaction,
\begin{equation*}
\begin{small}
    P \big( (u, v, t) \in E\mid u, \mathbf{x}_t \big) = \sum_{s \in \{ +1, -1 \}}
    P \big( s \mid u, \mathbf{x}_t \big)
    P \big( (u, v, t) \in E\mid u, \mathbf{x}_t, s \big).
\end{small}
\end{equation*}
Therefore, by using the definition from Equation~\ref{eq:link-probability}, the complete likelihood of interactions is

\begin{equation}
P(E\mid\mathbf{x}) = \prod_{(u, v, t) \in E}
            \sum_{s \in \{ -1, +1 \}}
            P(s)
            \frac{\kappa^s(x_{t, u}, x_{t, v})}
                 {\sum_{v' \in V^*_t} \kappa^s(x_{t, u}, x_{t, v'}) },
\end{equation}
where $P(s)$ is $\alpha_t$ for $s = 1$ and $(1 - \alpha_t)$ otherwise.

Similarly, the action likelihood is

\begin{equation}
  P(F\mid\mathbf{x}, \mathbf{w}, \mathbf{\sigma}) = \prod_{(v, a, t) \in F}
                    \frac{\kappa_\sigma(x_{t, v}, w_a)}
                         {\sum_{a' \in A} \kappa_\sigma(x_{t, v}, w_a') }
\end{equation}
by virtue of the probability defined in Equation~\ref{eq:feature-probability}.

We can use recursive Equation~\ref{eq:opinion-update} to substitute each occurence of $x_t$ in these formulas with a deterministic function of $x_0$ and $\signfun$.
Therefore, instead of writing $P(E \mid \mathbf{x})$, we write $P(E\mid\mathbf{x}_0, \signfun)$.
The details of this function are explored in Appendix~\ref{sec:reducing-to-x0}.

Now, we wish to maximize the log likelihood with respect to the latent variables ${\Theta
= (\mathbf{x}_0, \signfun, w, \sigma)}$ given the observed ones ${\Omega = (E, F)}$:

\begin{equation}
  \widehat{\Theta} =
  \argmax_{\Theta} \log P(E \mid \Theta) + \log P(F \mid \Theta) \\
\end{equation}

Optimizing this function is not straightforward as the expression for the latent variables $\Theta$ contains $\signfun$ --the sign of each arc in the interaction graph-- which is a discrete variable, thus leading to an integer programming problem.
We cannot solve this problem via a standard linear relaxation of the sign, since it would mean to define cases ``in between'' acceptance and contrast.
Such an approximation would defeat our purpose of translating a classic opinion dynamics model as faithfully as possible.

We therefore choose to employ the expectation-maximization (EM) technique.
In addition, to make the problem tractable, we resort to incremental learning approach in designing the algorithm.

\subsection{Online EM}
To apply EM, we choose a set of parameters $\theta = (\mathbf{x}_0, \mathbf{w}, \boldsymbol{\sigma})$ from our latent variables $\Theta$.
We thus wish to maximize the joint distribution $P(E, F \mid \mathbf{x}_0, \mathbf{w}, \boldsymbol{\sigma})$ given observed variables $\Omega = (E, F)$, the latent variables $\signfun$, and the parameters $\theta$.
Recall that solving this problem requires finding an assignment of the latent variables $\signfun$ such that for every observed arc $(u, v, t) \in E$ we have a sign $s \in \{-1, +1\}$.
Alas, this formulation would require the summation of the M step to consider all possible ${\signfun: E \rightarrow \{ -1, +1\}}$, which are $2^{|E|}$.

To simplify this problem, let us consider our process as an online task.
At each time step, our algorithm is presented  with the new interactions $E_t$.
The algorithm needs to decide their sign, i.e., whether each interaction is positive or negative.
Then, it needs to update its estimate for the opinions of the actors accordingly.
While solving the assignment problem for interactions in time step $\tau$,
the algorithm can therefore consider interactions and actions exclusively from the past time steps $t \leq \tau$.

Formally, let us consider a time step $\tau$.
Then, ${E_{\tau} = \{(\cdot, \cdot, \tau) \in E\}}$ are the actor-actor arcs at time $\tau$ and $F_{\tau}$ are the actor-actions arcs.
Similarly, ${\signfun_{\tau}: E_{\tau} \rightarrow \{ -1, +1 \}}$ are the signs of the interactions at the same time $\tau$.
Let also $\signfun_{< \tau} := \bigcup_{t < \tau} \signfun_t$.
We wish for our algorithm to take inputs $( E_\tau, F_\tau, \signfun_{< \tau} )$ together with a previous estimate for $\widehat{\theta} = (\mathbf{x}_0, \mathbf{w}, \boldsymbol{\sigma})$, and to return as output the maximum a posteriori estimate for $\signfun_\tau$ and $\theta$.
The probabilities of all signs and of the presence of all the links are conditionally independent:
$
  \signfun(u, v, \tau ) \indep
  \signfun(u', v', \tau )
  \mid (\theta, \signfun_{< \tau})
$ and $
  \big( (u, v, \tau) \in E_\tau \big) \indep
  \big( (u', v', \tau) \in E_\tau \big)
  \mid (\theta, \signfun_{< \tau})
$. As a consequence, we can express the likelihood of the signs as a product of independent likelihoods \begin{equation}
\label{eq:independent-assumption}
  P ( \signfun_\tau \mid E_\tau,F_\tau,\theta ) =
  \prod_{u, v : (u, v, \tau) \in E}
    P ( \signfun(u, v, \tau) \mid E_\tau,F_\tau,\theta ),
\end{equation}
which allows the algorithm to treat each separately.
Note that this result requires the online assumption.
Without it, since the opinions $\mathbf{x}_{t + 1}$ depend on $\signfun_{t}$, in the general case $ {
  \signfun(u, v, t ) \nindep \signfun(u', v', t' )
} $.

Therefore, thanks to the online assumption and Equation~\ref{eq:independent-assumption}, we can define the following expectation-maximization steps:

\medskip

\begin{mdframed}[backgroundcolor=gray!10,roundcorner=10pt]
\begin{squishlist}
  \item \emph{E Step}. For each arc $(u, v, \tau) \in E_\tau$ we evaluate
\begin{small}
  \begin{multline}
    \label{eq:e-step}
    q_{s, u, \tau} := P( s \mid (u, v, \tau) \in E_{\tau}, \widehat{\theta}) \\
    = \frac{ P( (u, v, \tau) \in E_{\tau} \mid u, s, \widehat{\theta}) \cdot P ( s \mid u, \widehat{\theta})}
           { \sum_{s' \in \{-1, +1 \}} P( (u, v, \tau) \in E_{\tau} \mid u, s', \widehat{\theta} )}
  \end{multline}
\end{small}
 where
\begin{small}
  \begin{equation*}
    P(s \mid u, \widehat{\theta})=\begin{cases}
    \alpha_t & \text{if } s = 1 \\
    1 - \alpha_t & \text{otherwise}
   \end{cases}
 \end{equation*} \end{small}
 and $P( (u, v, \tau) \in E_{\tau} \mid u, s, \widehat{\theta})$ is defined as in Equation~\ref{eq:link-probability}.

  \item \emph{M Step}. We update parameters $\theta$ in order to increase the following function
  $\mathcal{Q}(\theta)$:
  \begin{small}
  \begin{multline}  \label{eq:m-step}
    \sum\limits_{(u, v, \tau) \in E_\tau} \sum\limits_{s \in \{ -1, +1 \}} q_{s, u, \tau} \\
    \log \Big( P( (u, v, \tau) \in E_\tau\mid s, u, \theta)  P(F_{\tau, u}\mid\theta) \Big)
    = \log P(F_\tau\mid\theta) + \\
    \sum\limits_{(u, v, \tau) \in E_\tau} \sum\limits_{s \in \{ -1, +1 \}}
    q_{s, u, \tau} \log P( (u, v, \tau) \in E_{\tau} \mid s, u, \theta) \\
  \end{multline}
  \end{small}
  \text{where}
  \begin{small}
  \begin{equation}
  \label{eq:feature-likelihood}
  \log P(F_\tau\mid\theta) =
      \sum_{(v, a, \tau) \in F}
        \log \left( %
                    \frac{\kappa_\sigma(x_{\tau, v}, w_a)}
                         {\sum_{a' \in A} \kappa_\sigma(x_{\tau, v}, w_{a'}) }
            \right) %
  \end{equation}
  \end{small}
  and $P( (u, v, \tau) \in E_{\tau}, s \mid u, \theta)$ is again defined by Eq.~\ref{eq:link-probability}. %
 \end{squishlist}
\end{mdframed}

\smallskip

To increase the function in Eq. \ref{eq:m-step}, we employ gradient descent, and maximize it w.r.t. $\mathbf{x}_0, \mathbf{w}, \boldsymbol{\sigma}$.
The EM algorithm we have thus defined is summarized in Appendix~\ref{sec:pseudocode} (Algorithm~\ref{alg:online-step}).
It can be applied to one time step at a time, and considers only information coming from the previous time steps to update its parameters.
Starting from $t=0$, at each time step the algorithm is initialized with the current best estimate for its parameters, it updates them with new information, and emits the results for $\signfun_t$.
The resulting $\signfun_t$ and the updated parameters are then in turn used for the next time step estimate.
This schema is depicted in Figure~\ref{fig:online-chain} and summarized in Appendix~\ref{sec:pseudocode} (Algorithm~\ref{alg:complete-learning}).
This process can also be re-iterated: at each epoch, the whole learning process from $t=0$ to $t=T$ is repeated,
starting with the parameter estimates from the previous epoch, and continuously updating the parameters.
In practice, we repeat this process for a fixed number of epochs ($2$ in all the reported experiments).
Moreover, as common practice with EM algorithms, we employ a multiple restart approach: for each run, we repeat the learning process a number of times ($4$ in all the reported experiments) while changing the initial random seed; then, we pick the one with highest likelihood.

\begin{figure}
    \centering
    \includegraphics[width=.9\columnwidth]{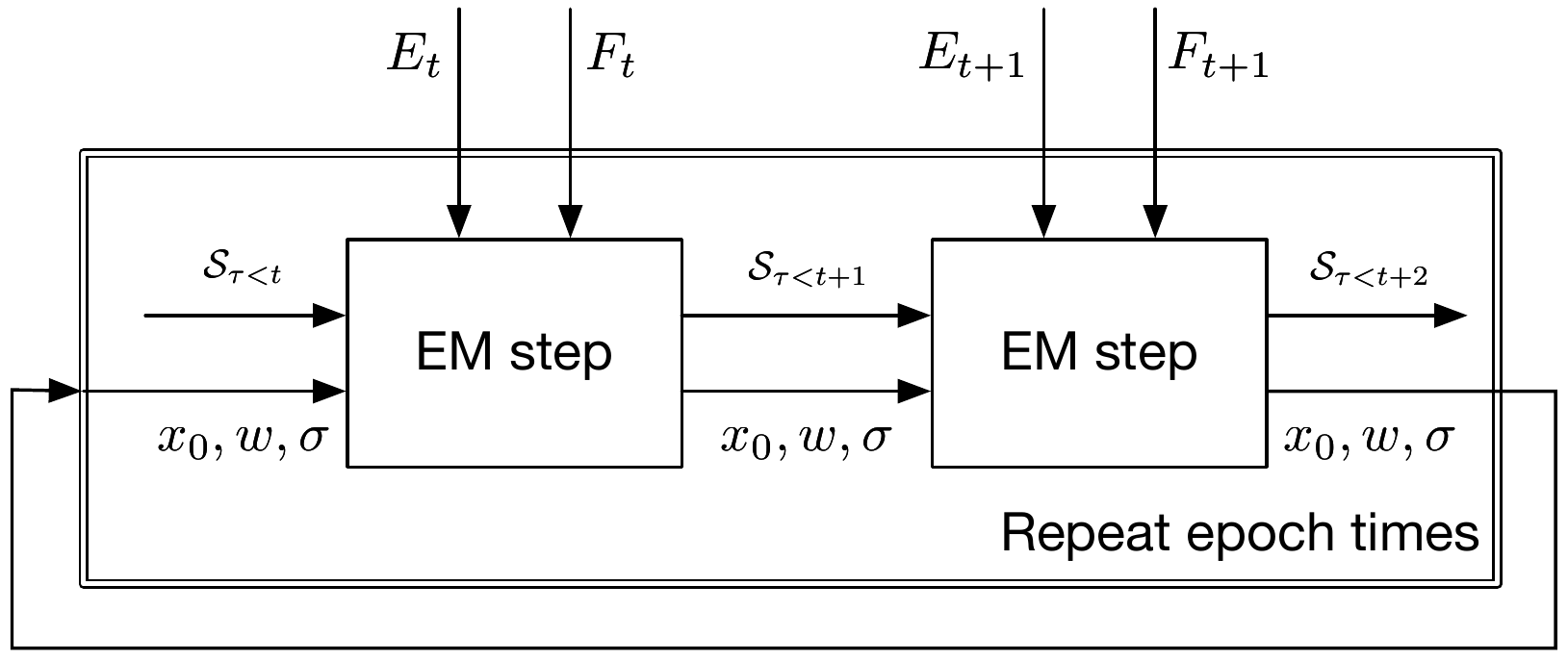}
    \vspace{-2mm}
    \caption{Schema of the proposed online learning process. Starting from $t=0$, for each time step $t$, the EM algorithm is presented with the estimate problem for the given time step; it updates the parameters and emits the estimate for $\signfun_t$. This whole process, from $t=0$ to $t=T$, can be repeated for a fixed number of epochs to improve the final estimates.}
    \label{fig:online-chain}
    \vspace{-4mm}
\end{figure}

What is the complexity of these computations? From Equations~\ref{eq:e-step}~and~\ref{eq:feature-likelihood}, it follows that the complexity for the E step is $\mathcal{O}(nm)$ where $n=|V^*_\tau|\leq|V|$ and $m=|E_\tau|$; for the M step, it is $\mathcal{O}(nm + n'm')$
where $n'=|A|$ and $m'=|F_\tau|$.%
Empirically, we report that running our framework on a common laptop for 2 epochs on $10$ time steps and $1\,000$ nodes, %
takes $140$ seconds for each restart; for $3\,000$ nodes, %
$1\,254$ seconds.

\section{Empirical Assessment}
\label{sec:experiments}

We focus on the following three research questions:
\begin{squishlist}
\item[\textbf{RQ1:}] Can we recover the micro parameters of the opinion dynamics model? (Section \ref{sec:rq1})
\item[\textbf{RQ2:}] Given a data trace from the generative process, can we find which macro-level scenario generated it?  (Section \ref{sec:rq2})
\item[\textbf{RQ3:}] Can the estimated parameters of the opinion dynamics model on real data explain real user behavior?  (Section \ref{sec:rq3})
\end{squishlist}
To answer these questions we use a mix of synthetic data and real data; the latter represent a 10-year data set we crawled from the social rating and discussion website Reddit.
While RQ1 and RQ2 deal with the internal validity of our proposal (inference algorithm and model selection framework, respectively),
RQ3 tests the external validity of the inferred model parameters.
The results of model selection on real data are quite interesting: Section~\ref{sec:discussion} discusses some possible interpretations.
We publicly release our implementation and data set to facilitate reproducibility.\footnote{\url{https://github.com/corradomonti/learnable-opinion-dynamics}}

\begin{table}
	\caption{Distance (mean average error) and classification accuracy (F1 and average precision) on synthetic experiments; we report average and std. dev. across $8$ generated data traces.}
	\vspace{-0.5\baselineskip}
	\begin{footnotesize}
		\begin{tabular}{lllll}
\toprule
{} & MAE $\mathbf{x}_0$ & MAE $\mathbf{w}$ & $\mathcal{S}$ F1-score &     $Z$ Av.Prec. \\
\midrule
Non-commitment  &    $0.13 \pm 0.02$ &  $0.18 \pm 0.03$ &        $0.99 \pm 0.02$ &  $0.95 \pm 0.01$ \\
Balanced        &    $0.16 \pm 0.04$ &  $0.14 \pm 0.01$ &        $1.00 \pm 0.00$ &  $0.96 \pm 0.02$ \\
High contrast   &    $0.13 \pm 0.02$ &  $0.16 \pm 0.03$ &        $0.98 \pm 0.01$ &  $0.97 \pm 0.01$ \\
High acceptance &    $0.34 \pm 0.16$ &  $0.26 \pm 0.12$ &        $0.90 \pm 0.08$ &  $0.93 \pm 0.03$ \\
\bottomrule
\end{tabular}

	\end{footnotesize}
	\label{tab:syndata}
	\vspace{-\baselineskip}
\end{table}

\subsection{Recovering opinion micro parameters}
\label{sec:rq1}

RQ1 deals with the micro parameters of our models: the opinions of agents and actions, and the signs of the interactions.
To test our inference algorithm, we generate synthetic data traces according to the model by~\citet{jager2005uniformity}.
The set of macro parameters ($\epsp,\epsn$) for the given trace, which define a \emph{scenario}, are taken from the same work, and reported in Figure~\ref{fig:example-scenarios}.

Each scenario represents different assumptions about the behavior of the actors.
A \emph{high acceptance} scenario is characterized by a high latitude of acceptance \epsp, which results in consensus among the actors.
Conversely, a \emph{high contrast} scenario, generated by a low latitude of contrast \epsn, results in frequent backfires and a polarized system.
A low \epsp and a high \epsn generate a scenario of \emph{non-commitment}, where the opinions are stable and fragmented.
Finally, in a \emph{balanced} scenario, the distance in opinion space is equally divided among acceptance, neutral, and contrast zones, and opinions cluster into a small number of attraction points.

Actions are not part of the original model, so we generate them according to the stochastic process described in Section~\ref{sec:generative-features} (initialized uniformly in $[-1,1]$).
For each scenario we generate $8$ different data traces.
Then, we fit the model with the set of macro parameters corresponding to the specified scenario.
Finally, we measure how close the inferred micro parameters (opinions and interaction signs) are to the generated ones.

Table~\ref{tab:syndata} shows four measures of the quality of our predictions in the four scenarios, on average across 8 experiments. First, we show the mean absolute error between the original $\mathbf{x}_0$ and its estimate,\footnote{Since the estimate is symmetric, we take the best between $\mathbf{x}_0$ and $-\mathbf{x}_0$.} and the same for the action opinions $\mathbf{w}$.
Then, we treat assigning a positive or a negative sign to each interaction in $G$ as a binary classification problem, and compute the F1 score with respect to the original signs.
Finally, we measure how well our model captures the actor-action graph by taking all its edges $F$, a sample of non-existing edges $(v, a, t) \notin F$ of the same cardinality $|F|$, and we compute the average precision of our model in separating the two.

In most scenarios, the inferred opinions are very close to the generated ones, the signs of the interactions are almost perfectly recovered, and the actions are well fit by the model.
The high-acceptance scenario proves to be more challenging, as the final consensus equilibrium blurs the individual opinion of each actor.

\newcommand{\drawdistances}[2]{%
\begin{tikzpicture}[thick,xshift = 10,scale=1.33, every node/.style={scale=0.8}]
			\draw[fill, blue!40!gray] (0,0) circle (1pt) node {};
			\draw[-., blue!40!gray, line width=2.]    (0,0) to (#1,0);
			\draw[-, gray]    (#1,0) to (#2,0);
			\draw[-, red!40!gray, line width=2.]    (#2,0) to (2,0);
			\draw[fill, red!40!gray] (2,0) circle (1pt) node {};
			
		  \draw[left=2mm] (0,0) node {0};
			\draw[right=2mm] (2,0) node {2};
			\draw[left=0.01mm, above=2mm] (#1+0.05,0) node {$\epsp=#1$};
			\draw[fill, blue!40!gray] (#1,0) circle (1pt) node {};
			\draw[right=0.01mm, below=2mm] (#2+0.05,0) node {$\epsn=#2$};
			\draw[fill, red!40!gray] (#2,0) circle (1pt) node {};
\end{tikzpicture} }

\begin{figure*}[t!]
\begin{center}
	\begin{tabular}{rrrr}
		\drawdistances{0.6}{1.2} &
		\drawdistances{0.4}{0.6} &
		\drawdistances{1.2}{1.6} &
		\drawdistances{0.2}{1.6} \\
		\hspace{-2mm}\includegraphics[width=0.24\linewidth]{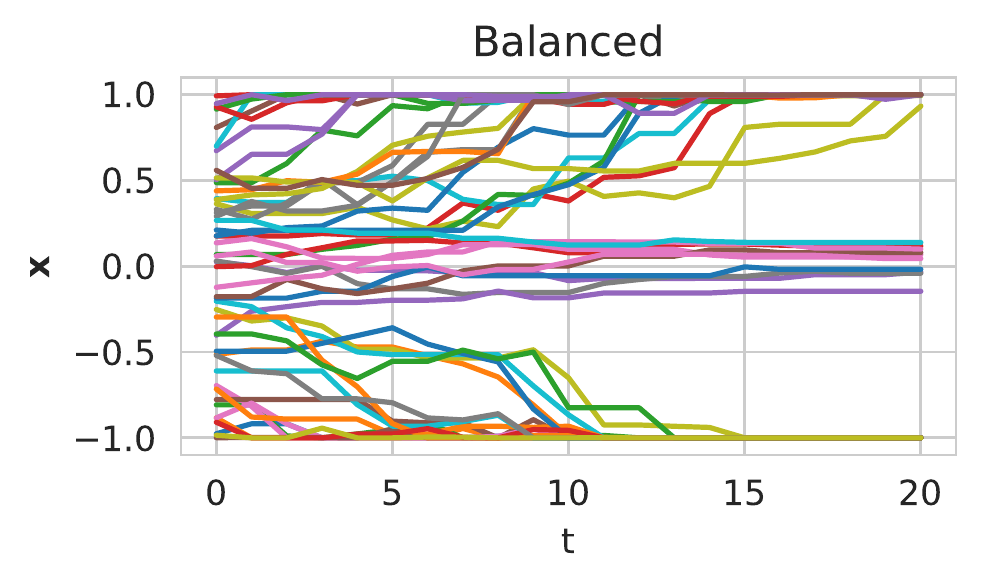}
		&
		\hspace{-2mm}\includegraphics[width=0.24\linewidth]{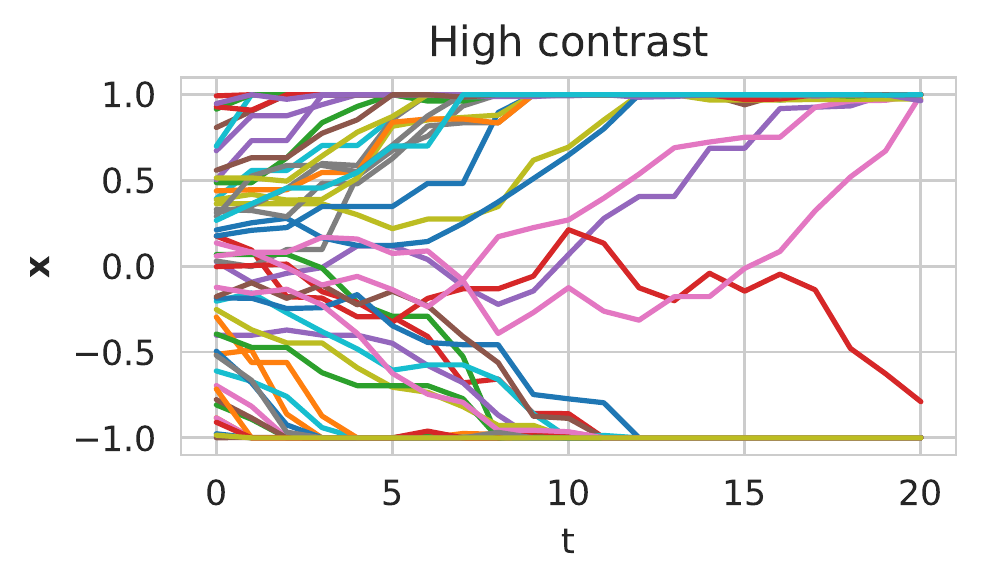}
		&
		\hspace{-2mm}\includegraphics[width=0.24\linewidth]{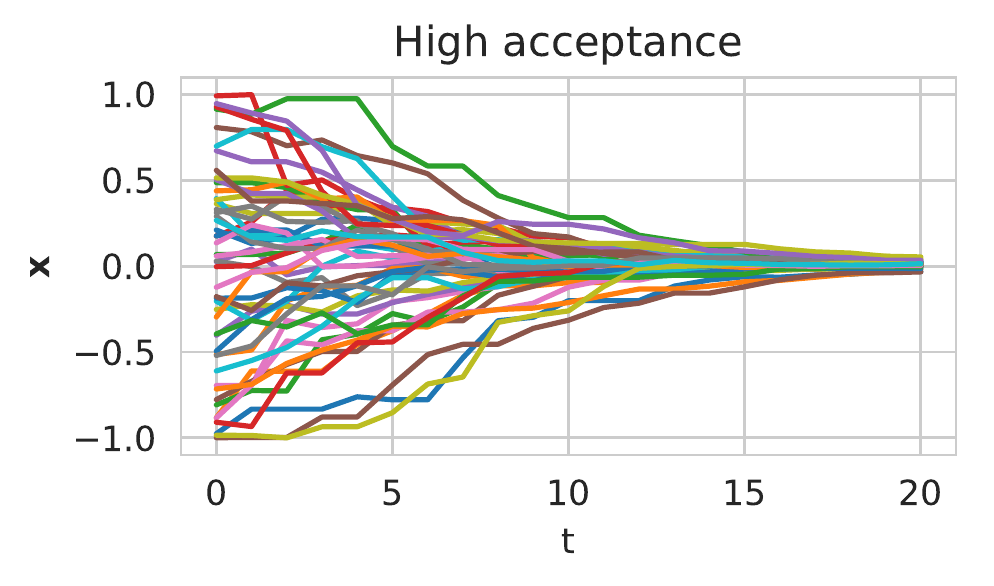}
		&
		\hspace{-2mm}\includegraphics[width=0.24\linewidth]{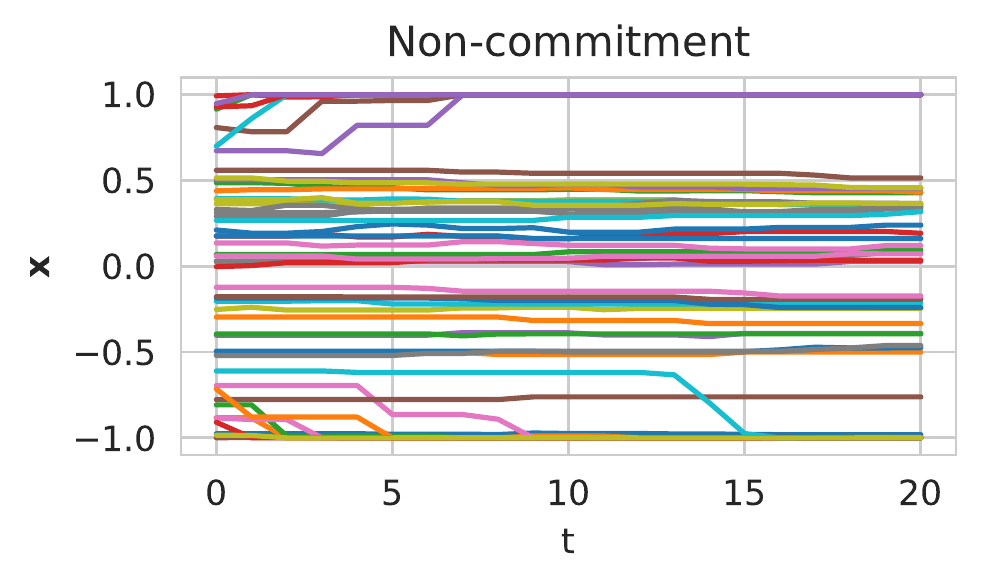}
	\end{tabular}
\vspace{-0.9\baselineskip}
\caption{ %
Examples of synthetic data traces generated in each scenario. %
Plots represent the opinion trajectories along time. }
\label{fig:example-scenarios}
\end{center}
\vspace{-0.5\baselineskip}
\end{figure*}
\newcommand{\hyperparameterstable}{Figure~\ref{fig:example-scenarios}}

\subsection{Discriminating macro-level scenarios}
\label{sec:rq2}
We now ask whether our framework is able to discriminate which scenario generated a given data trace.
If our framework can accomplish this task, we can use it to assess the plausibility of assumptions of opinion formation models by testing them on real data (as we show in Section~\ref{sec:rq3}).
Operationally, we run our algorithm against the data trace with different sets of macro parameters, one for each scenario hypothesis we wish to test.
Finally, we look at the likelihood obtained under each hypothesis.

\begin{figure*}[t!]
\begin{center}
		\includegraphics[width=.49\linewidth,valign=b]{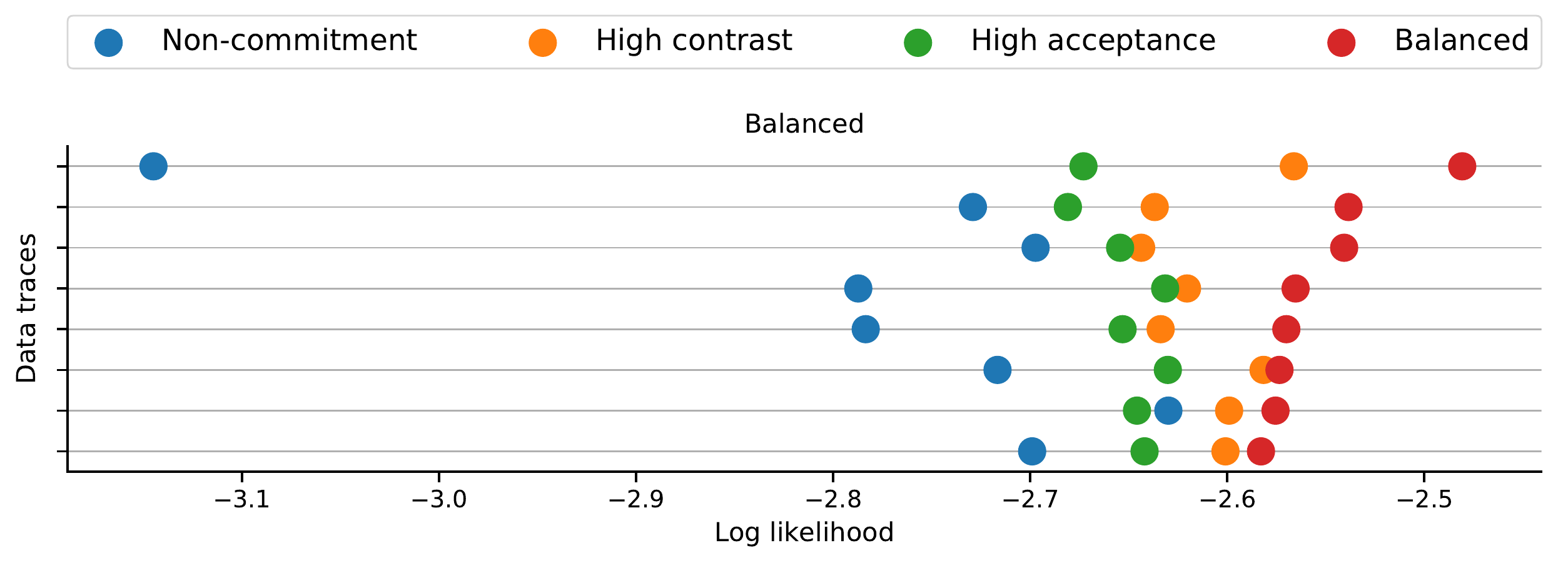}
		\includegraphics[width=.49\linewidth,valign=b]{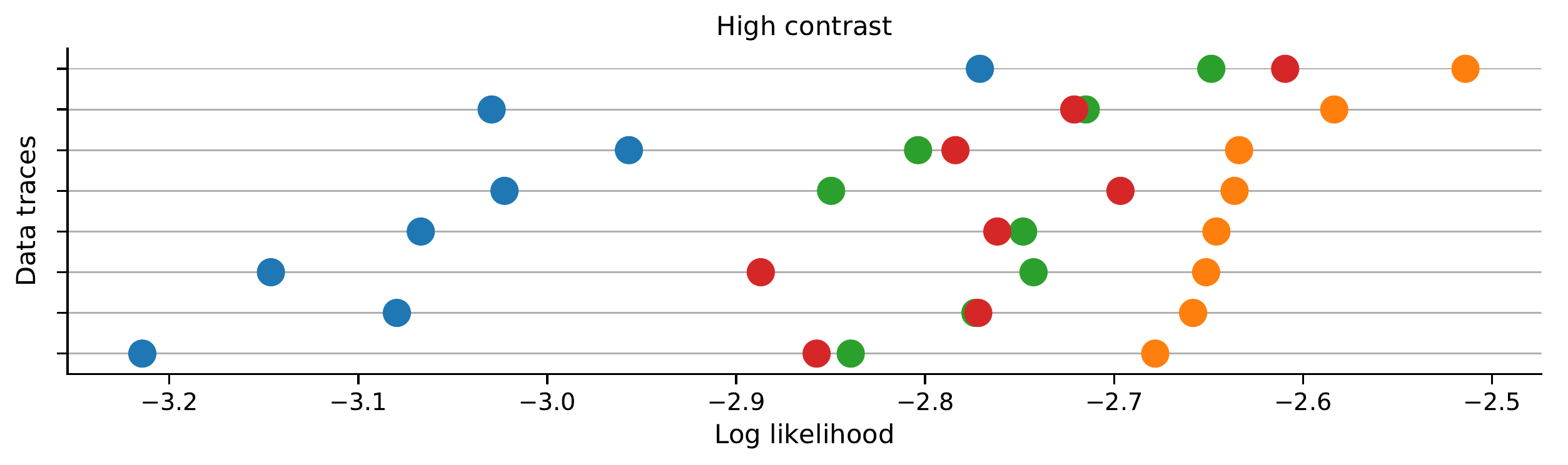} \\
		\includegraphics[width=.49\linewidth,valign=b]{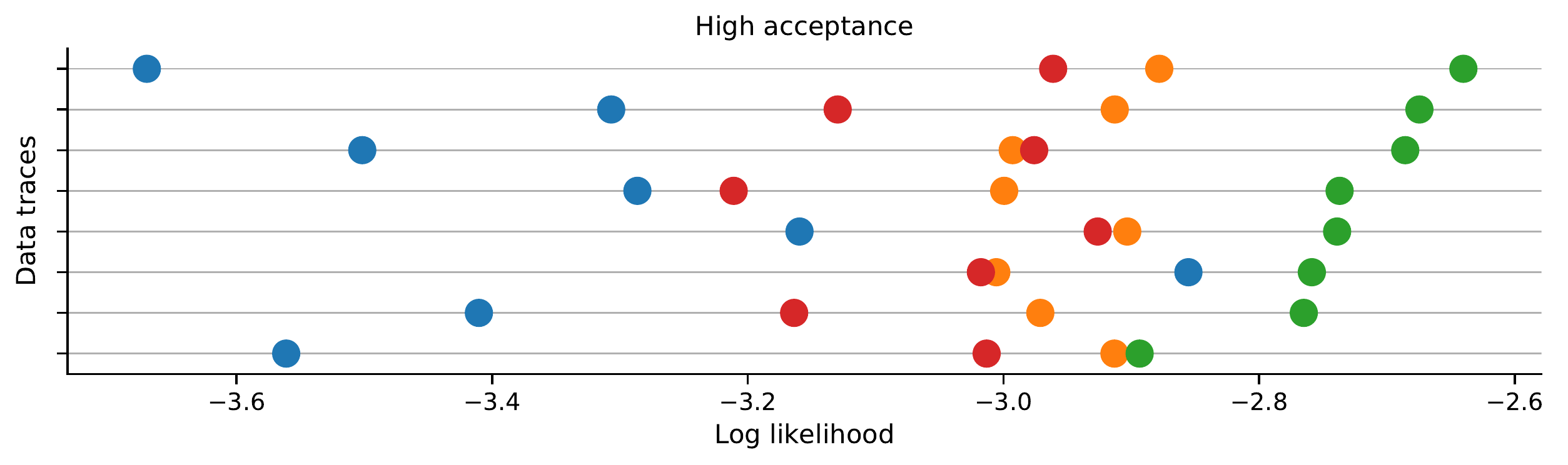}
		\includegraphics[width=.49\linewidth,valign=b]{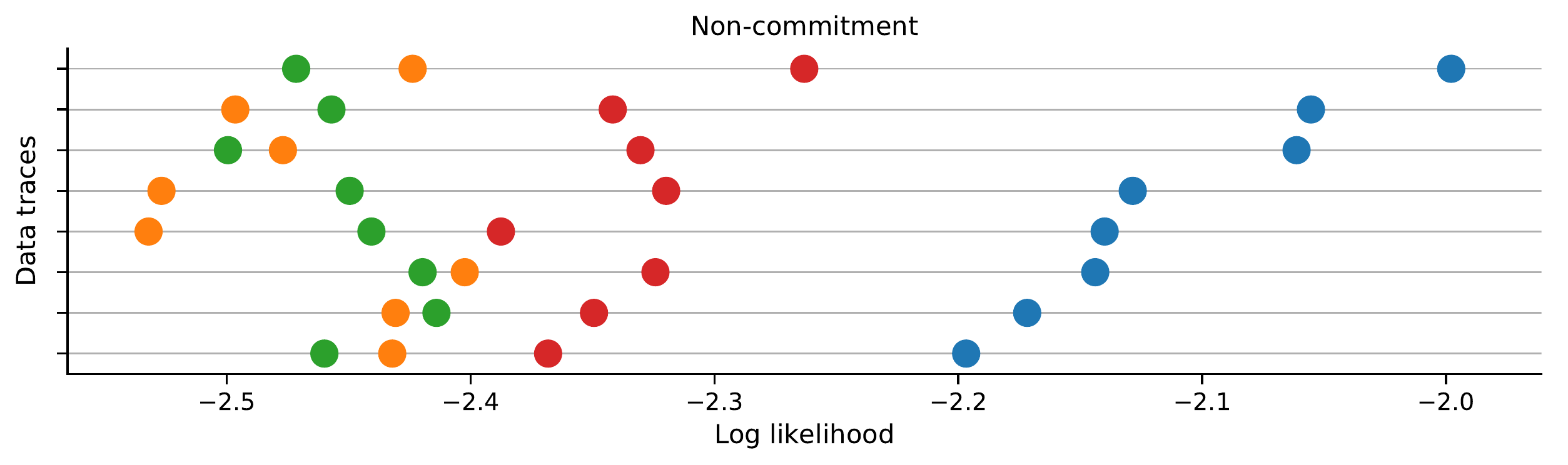}
\vspace{-\baselineskip}
\caption{Each panel represents a different macro parameters scenario, and each line on the Y axis a different data trace generated according to that scenario. On the X axis we report, for each data trace, the log likelihood obtained by the the best estimate of our generative model, initialized with a given set of macro parameters, one per color. Rightmost corresponds to highest likelihoods. We see that the estimate corresponding to the true macro parameters has the highest likelihood.}
\label{fig:synth-hyper-params}
\end{center}
\vspace{-\baselineskip}
\end{figure*}

In our experiments, we generate $8$ synthetic data traces for each scenario.
Then, for each data trace, we run our algorithm with the four different macro parameters encoding each scenario hypothesis.

Figure~\ref{fig:synth-hyper-params} shows the likelihood of each different generated data trace under each tested scenario.
In all cases, the most likely set of macro parameters found by our framework is the true one that generated the data trace itself.
Specifically, it is close to a perfect accuracy for every scenario except ``high-acceptance'', for which the results are still mostly positive.

\subsection{Opinion dynamics on real data}
\label{sec:rq3}
In this section we apply the framework to real data from Reddit to explore the prominence of the backfire effect, i.e.,
to see whether a scenario with large latitude of contrast is likely.

\spara{Dataset.}
We gather Reddit data from 2008 to 2017, and bucket it so that one time step corresponds to a month (120 time steps in total).
Reddit users are actors, while posting in a subreddit corresponds to an action.
User $v$ replying in a comment thread to user $u$ at time step $t$ corresponds to an interaction $(u, v, t)$.
We sample from both users and subreddits to create our dataset.
In order to study US political discussion, we choose \texttt{r/politics} as our seed subreddit
and pick the $50$ most similar subreddits to \texttt{r/politics} according to cosine similarity over a vector representation of the subreddits based on latent semantic analysis, which captures subreddits whose user base is similar to the seed one.\footnote{\url{https://www.shorttails.io/interactive-map-of-reddit-and-subreddit-similarity-calculator}}
Resulting subreddits include political ones such as \texttt{r/democrats}, ones dedicated to specific politicians such as \texttt{r/hillaryclinton} and \texttt{r/The\_Donald}, and ideological ones such as \texttt{r/Libertarian}.
We then sample users posting a minimum of $10$ comments per month on \texttt{r/politics} for at least half of the months, which gives us $375$ users.
The resulting action graph has approximatively $144$k actor-action arcs, while the interaction graph has approximatively $90$k actor-actor arcs.%

Reddit allows to up/down-vote posts, which represents the social feedback of the community.
The score of a post on a subreddit is a function of the up- and down-votes received by it from other users in that subreddit.
It represents how well-received the post is by the specific subreddit community.
A negative score means that the post has been disapproved by the community, possibly because it expresses a point of view that is far from the norm of the subreddit.
A high absolute score indicates a high attention for the post, i.e., it has been read and voted by a large number of users in the subreddit.

We consider two different application settings for the framework: with or without an \emph{anchored axis}.
An anchored axis refers to fixing the position in opinion space of a set of actions.
In particular, we fix two actions as the extremes of the opinion space.
This way, we create an axis along which all other actions (and actors) lay.
By changing the definition of the axis, we can explore different semantics for the latent opinion space.

In the experiments, we explore two anchors for the axis:
one between \texttt{r/democrats} or \texttt{r/Republican}, by fixing their latent opinion point to $w_a = \{-1, 1\}$, respectively,
and one with respect to \texttt{r/The\_Donald}, by fixing its latent opinion point to $w_a = 1$.
The first anchoring represents the traditional political spectrum in US, the second one represents the closeness to Donald Trump supporters.
In summary, we have three different axes: a free one (\texttt{None}), a left-right one (\texttt{r/democrats} --- \texttt{r/Republican}), and a unipolar one (\texttt{r/The\_Donald}).
For each of the two possible cases w.r.t. anchored axis, we test a set of parameters as the ones presented in \hyperparameterstable.

To verify that the model is capturing the underlying behavior from the data, we employ two external validation metrics.
These metrics are completely hidden from the framework, and try to capture the user behavior on Reddit:
\begin{squishlist}
\item \textbf{User-subreddit score:}
	For each user, subreddit, and time step, we compute the average score of all the posts the user has submitted to the given subreddit in the specific month. This score is a proxy for how well-received the opinions of the user are in the specific subreddit.
\item \textbf{User-user conflicts:}
	We identify set of user-user interactions that exhibit conflictual behavior. The intuition is that when a reply to a positive-score comment has a negative score (or vice versa), the two authors are probably expressing conflicting points of view. To capture this behavior, we define a conflictual interaction of comment $x$ to comment $y$ when $x$ and $y$ have scores with opposite signs. We restrict our attention to comments that have attracted some attention in the community, i.e., with a minimum absolute score of $10$.
\end{squishlist}
We now present results for these two external evaluation metrics by using the best estimate according to our \emph{internal} validation metric, the log likelihood.
The best-fitting scenario is a \emph{high-acceptance} one anchored on \texttt{r/The\_Donald}, but results are qualitatively similar for a \emph{non-commit} scenario.
For this experiment, the average precision on the real user-subreddit links $F$ (as defined in Section~\ref{sec:rq1}) is $0.934$.
We measure the Pearson correlation coefficient between the user-subreddit score and the distance between user and subreddit in opinion space, as inferred by our model.
Our hypothesis is that a higher score corresponds to a lower distance between the two, and therefore the correlation should be negative.
This behavior is consistent with the idea that opinions close to the norm of the subreddit are the ones that get the most appreciation, as can be explained by cognitive dissonance theory~\citep{festinger1957theory}.

\begin{figure}[t]
\begin{center}
\includegraphics[width=\linewidth]{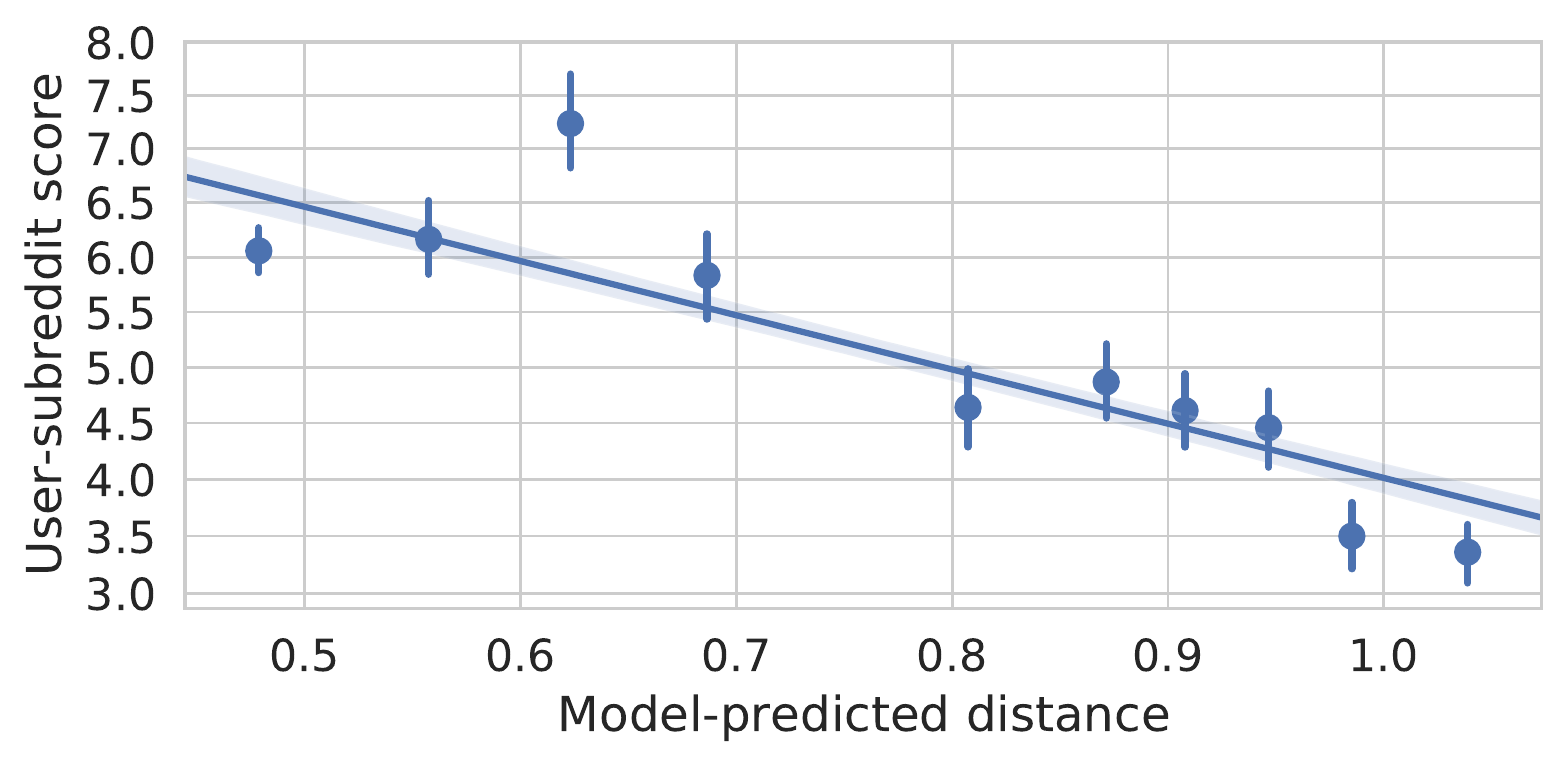}
\vspace{-1.5\baselineskip}
\caption{Univariate regression between user-subreddit score and distance between user and subreddit in opinion space as inferred by our model. The correlation coefficient is negative with a value of $-0.127$ and highly significant ($p < 10^{-6}$). Error bars represent the $95\%$ confidence intervals.
}
\label{fig:regression-distance-score}
\end{center}
\vspace{-\baselineskip}
\end{figure}

\begin{figure}[t]
\begin{center}
	\begin{tabular}{cc}
		\hspace{-4mm}
		\includegraphics[height=3cm]{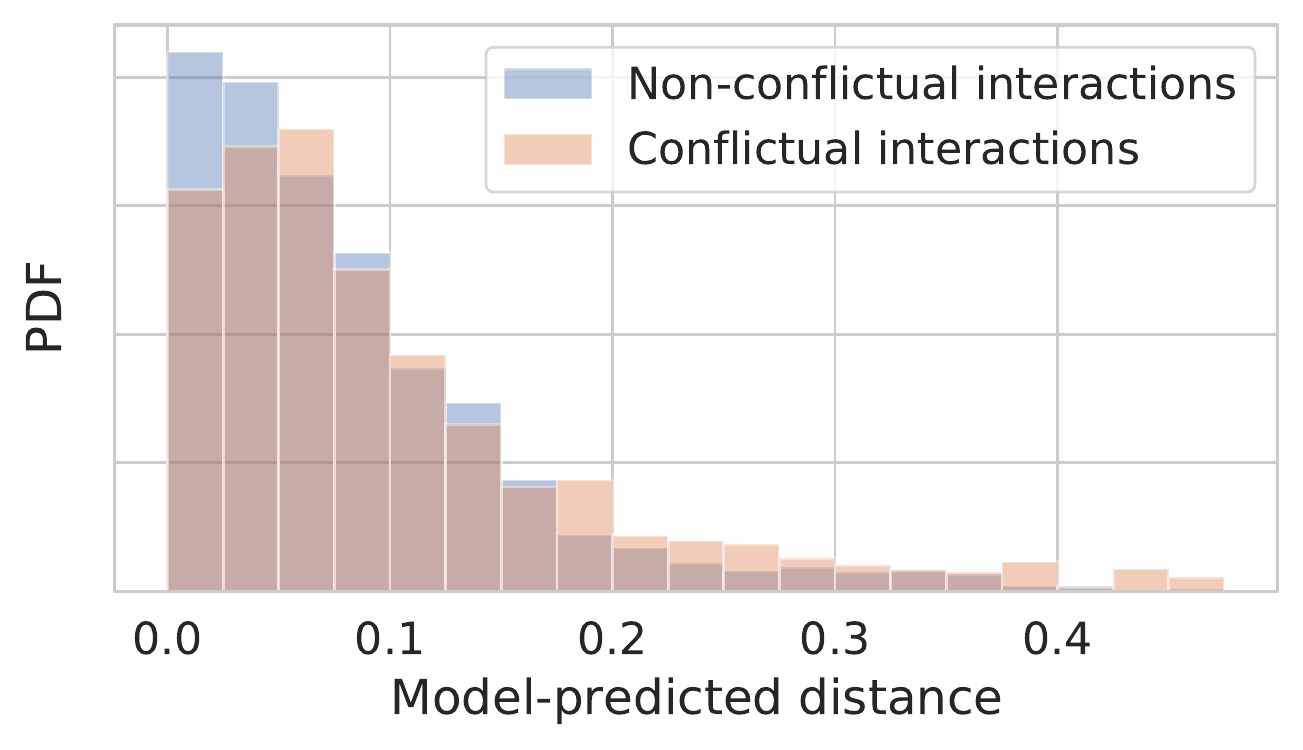}
		&
		\includegraphics[height=3.1cm]{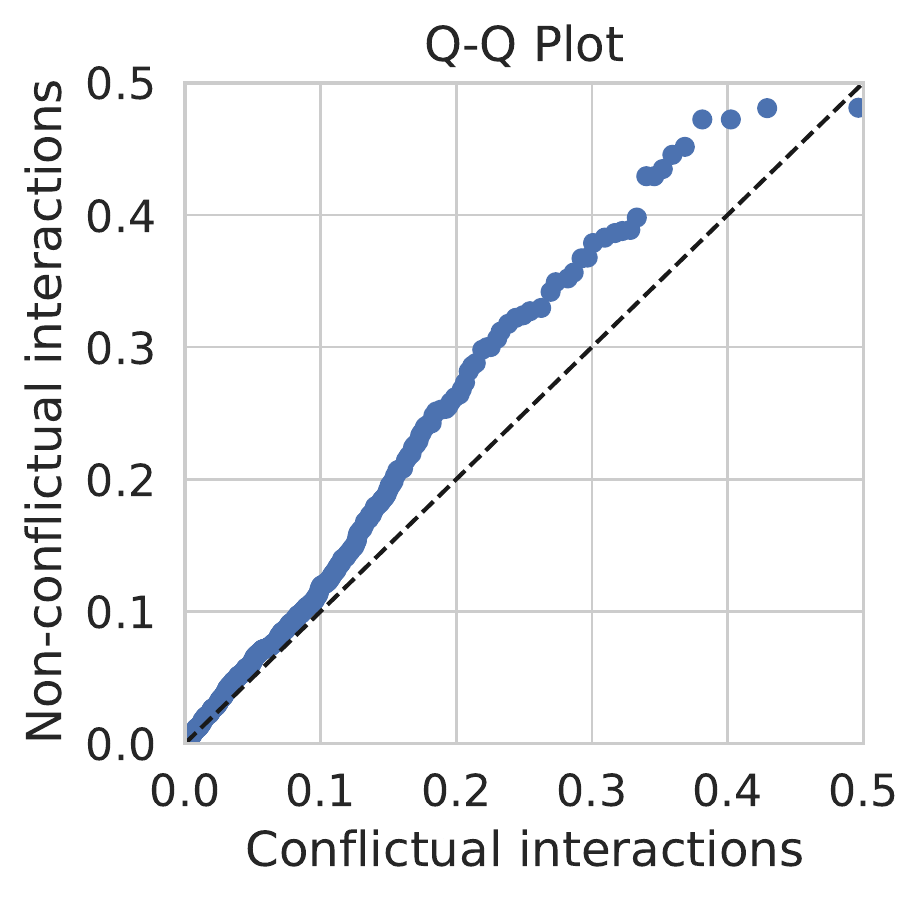}
		\vspace{1mm}
	\end{tabular}
\vspace{-\baselineskip}
\caption{Distributions of distances in opinion space between users during conflictual interactions (orange) and non-conflictual interactions (blue). Conflictual interactions have a significantly larger distance on average ($p < 10^{-6}$).}
\label{fig:conflict-distance-distribution}
\end{center}
\vspace{-\baselineskip}
\end{figure}

\begin{figure}[tbp]
\begin{center}
\includegraphics[width=.8\linewidth]{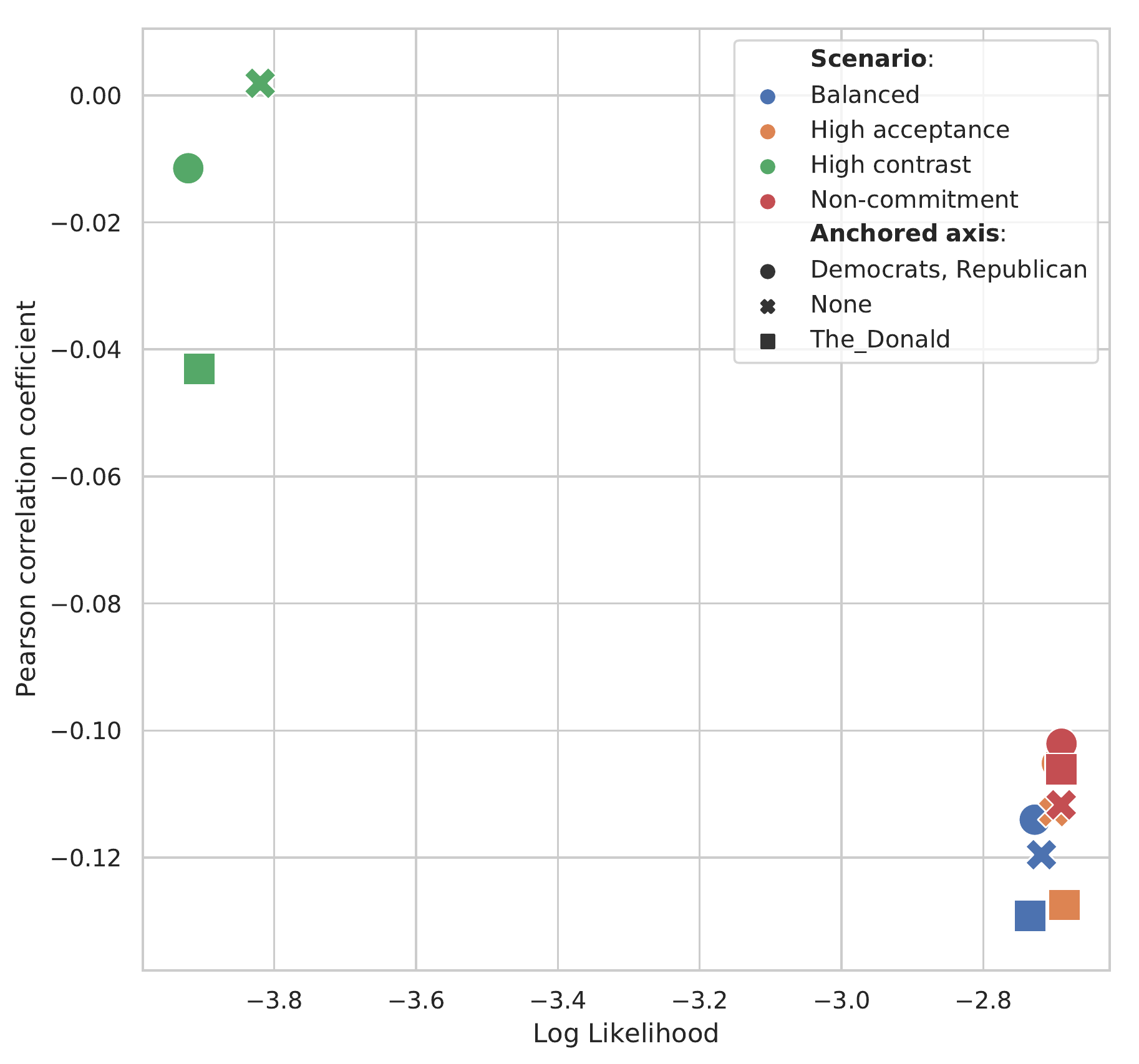}
\vspace{-\baselineskip}
\caption{Scatter plot of the results obtained by different hyper-parameters on estimates on Reddit data, with respect to an external and an internal evaluation metric. On the X axis, we report the internal objective function of the model, its log likelihood. On the Y axis, the external evaluation metric: the Pearson correlation between the user-subreddit latent opinion distance and the average user-subreddit scores (bottom right is better).}
\label{fig:likelihood-correlation}
\end{center}
\vspace{-\baselineskip}
\end{figure}

Figure~\ref{fig:regression-distance-score} shows the regression of the user-subreddit score as a function of their inferred distance.
The relationship between the variables is negative as predicted by our hypothesis.
In other words, users that are more popular within a community are the closest to that community in our opinion space.
This result confirms that the parameters inferred by our model, in particular the opinions of the users and the subreddits, capture some of the drivers behind user voting behavior.

For the second validation metric, for each interaction, we measure the distances between pairs of users in opinion space at the time of the interaction.
We compute these distances for the interactions specified above, and also for a control group of non-conflictual interactions (i.e., both scores are positive).
We also apply the same minimum score threshold as above to select the non-conflictual interactions.
Our hypothesis is that conflictual interactions are more likely to happen between users that are further apart in opinion space.

Figure~\ref{fig:conflict-distance-distribution} shows the distributions for both kind of interactions.
The conflictual interactions present a higher average distance than the control group, which is consistent with our hypothesis.
A one-sided non-parametric Mann-Whitney U test confirms the hypothesis that a randomly selected conflictual interactions has a larger distance than a non-conflicting one ($p<10^{-6}$).
The median distances differ by $0.06$.
This result shows that our model is able to capture some of the mechanisms behind the emergence of conflicts.
The small effect size is to be expected, as conflicts might happen for a number of reasons not directly related to the ideological positions of the users interacting.
Nevertheless, the strong statistical association between the parameters inferred by our model and real-world user behavior as measured from noisy data is a clear signal that our algorithm is able to capture some latent user characteristic.

Finally, we explore the selection of the macro parameters from \hyperparameterstable~with respect to the external validation metrics.
We use the user-subreddit score as it is numerical, and thus can offer a higher granularity for better presentation.
Figure~\ref{fig:likelihood-correlation} shows the relationship between the likelihood of the model given the set of macro parameters, and the correlation of user-subreddit scores with user-subreddit distances (model parameters).
As such, this graph shows the relationship between an internal evaluation metric (the likelihood) and an external validation one (the score-distance correlation coefficient).
The two metrics agree for the most part, thus suggesting that we can use the likelihood to identify the most fitting model that explains real-world behaviors.

\section{Discussion and future work}
\label{sec:discussion}

We have proposed \nameacr: a learnable generalization of an opinion dynamics model.
It retains the explainability and causal interpretation of agent-based models, by describing the underlying data generation process via latent and observed stochastic variables.
We have shown how to cast a classic agent-based opinion dynamics model into our framework, and designed an algorithm infer its parameters from data.
Clearly, this model is a proof-of-concept, and a similar process can be applied to other opinion dynamics models to make them testable and learnable.
Since our work is based on a generalization of BCM, it should be easily applicable to other BCM extensions~\citep{delvicario2017modeling, quattrociocchi2011opinions}.
Thanks to recent efforts in unifying different opinion dynamics model under a common formalism~\citep{coates2018unified}, it might be possible to build \emph{general} learnable opinion dynamics model.
This framework could leverage social traces to validate empirically several assumptions on opinion dynamics, with the final goal of improving our understanding of how the human mind shapes ideas through social interactions.

Our experiments have shown that the framework is able to learn the micro-level parameters of the single actors.
For instance, we are able to distinguish positive interaction from negative ones.
We are also able to recover the latent opinion of actors, and their trajectory in time.
This feature allows fine-grained analysis of real individuals with the same techniques used to describe opinion dynamic models.
In other words, we are able to empirically quantify and verify the assumptions of opinion dynamics model at an individual level.

Moreover, we have shown the capabilities of our proposal for model selection.
The framework is able to identify the correct scenario (i.e., the set of macro-level parameters that encode the interaction rules) that generated a given data trace in synthetic experiments.
This capability is extremely useful for testing sociological assumptions, which can still be expressed as deterministic update rules for agents' internal states.

We have applied our framework to a real-world dataset coming from Reddit, and have shown that the best-fitting model is able to explain user-level behavior.
In particular, we are able to explain a trend in voting behavior of users on subreddits by looking at the learned micro parameters (the opinions of users and subreddits).

By using our framework for model selection on Reddit data, we find that the ``high acceptance'' and ``non-commitment'' scenarios are the most likely, and the ``high contrast'' one is the least likely by far.
Our model thus rejects the presence of a low latitude of contrast.
These results suggest that the backfire effect is negligible among active participants in Reddit's political conversation.

A possible explanation for our results is that a community such as Reddit, over a time span of a decade, tends to evolve more according to a consensus-creation mechanism than an internal polarization one.
For example, the social feedback inherent in the platform may stifle extreme opinions, and create more pressure towards mainstream attitudes.
Following new trends might be more appealing than the rejection created by polarization mechanisms.

\begin{table}[t]
\footnotesize
\centering
\caption{Position in latent opinion space of the top-20 most popular subreddits in our data. See Section~\ref{sec:discussion} for a discussion.}
\vspace{-0.5\baselineskip}
\begin{minipage}{.49\linewidth} %
\begin{tabular}{lrr}
\toprule
                      Subreddit &  $w_a$ &  $\sigma_a$ \\
\midrule
         \texttt{r/The\_Donald} &   1.00 &        0.69 \\
          \texttt{r/Republican} &   1.00 &        0.38 \\
         \texttt{r/progressive} &   0.99 &        0.58 \\
           \texttt{r/Economics} &   0.89 &        0.65 \\
         \texttt{r/Libertarian} &   0.88 &        0.61 \\
          \texttt{r/TrueReddit} &   0.87 &        0.60 \\
          \texttt{r/Futurology} &   0.84 &        0.67 \\
          \texttt{r/conspiracy} &   0.84 &        0.60 \\
                \texttt{r/news} &   0.52 &        0.61 \\
            \texttt{r/politics} &  -0.34 &        0.60 \\
\bottomrule                    %
\end{tabular}
\end{minipage}%
\begin{minipage}{.49\linewidth}
\begin{tabular}{lrr}
\toprule
                      Subreddit &  $w_a$ &  $\sigma_a$ \\
\midrule
           \texttt{r/worldnews} &  -0.53 &        0.59 \\
       \texttt{r/todayilearned} &  -0.65 &        0.60 \\
             \texttt{r/atheism} &  -0.83 &        0.60 \\
     \texttt{r/EnoughTrumpSpam} &  -0.84 &        0.53 \\
 \texttt{r/SandersForPresident} &  -0.89 &        0.55 \\
 \texttt{r/PoliticalDiscussion} &  -0.91 &        0.59 \\
       \texttt{r/worldpolitics} &  -0.95 &        0.62 \\
        \texttt{r/changemyview} &  -0.97 &        0.57 \\
        \texttt{r/Conservative} &  -1.00 &        0.47 \\
             \texttt{r/economy} &  -1.00 &        0.52 \\
\bottomrule
\end{tabular}
\end{minipage}                %
\label{tab:subreddit-opinion}
\vspace{-\baselineskip}
\end{table}

The proposed framework has numerous possible applications.
As an example, Table~\ref{tab:subreddit-opinion} reports the inferred positions in opinion space for the top-20 most popular subreddits in our dataset.
Here the latent opinion space is anchored so that \texttt{r/The\_Donald} (a community of Donald Trump supporters) is fixed at one extreme ($1.0$).
The position of many subreddit in opinion space seems reasonable and follows intuition.
The community of Bernie Sanders supporters (\texttt{r/SandersForPresident}) is correctly positioned near the other end of the spectrum.
A conspiracy group (\texttt{r/conspiracy}), which has been described as taking ``a pro-Trump bent'',\footnote{https://www.vox.com/2018/8/8/17657800/qanon-reddit-conspiracy-data} is placed very close to Donald Trump supporters.
This example shows how our model could be used to analyze opinion trajectories estimated under a specific set of hypothesis.

\bibliographystyle{ACM-Reference-Format}
\bibliography{references}

\clearpage
\appendix
\begin{footnotesize}

\begin{algorithm}[h!]
    \caption{EM Step for time step $t$}
    \label{alg:online-step}
    \begin{flushleft}
    \algorithmicrequire \; Graph $G_t = (V, E_t)$; actions $Z_t=(V, A, F_t)$;  $\mathcal{U}^{t}(\mathbf{x}_0)$; $\alpha_t$. \\
    \algorithmicensure \; opinions $\mathbf{x}_0$, actions $( \mathbf{w}, \mathbf{\sigma}) $, signs $\signfun_t: E_t \rightarrow \{-1, +1\}$
    \begin{algorithmic}[1]
            \State If not given, initialize $\mathbf{x}_0, \mathbf{w}, \mathbf{\sigma}$

            \State $V^*_t := \{ v \in V : (\cdot, v, t) \in E_t \}$

            \Repeat
                 \Comment{E Step}
                \State Compute $\mathbf{x}_{t} := \mathcal{U}^{t}(\mathbf{x}_0)$
                \For{$(u, v, t) \in E_t$}
                    \State $
                          p^+(u, v) := { \kappa^+(x_{t, u}, x_{t, v})} /
                                       {\sum_{v' \in V^*_t} \kappa^+(x_{t, u}, x_{t, v'}) }
                    $
                    \State $
                          p^-(u, v) := { \kappa^-(x_{t, u}, x_{t, v})} /
                                       {\sum_{v' \in V^*_t} \kappa^-(x_{t, u}, x_{t, v'}) }
                    $
                    \State \begin{equation*}
                          q^+(u, v) :=  \frac{ \alpha_t \cdot p^+(u, v) }
                                             { p^+(u, v) + p^-(u, v) }
                    \end{equation*}
                    \Comment{Eq.~\ref{eq:e-step}}
                    \State \begin{equation*}
                          q^-(u, v) := \frac{ (1 - \alpha_t) \cdot p^-(u, v) }
                                             { p^+(u, v) + p^-(u, v) }
                    \end{equation*}
                \EndFor

               \Comment{M Step}
                  \State Using $\mathbf{x}_{t} = \mathcal{U}^{t}(\mathbf{x}_0) $, do:
                  \State \phantom{Using..} Update $\mathbf{x}_0$ by ascending the gradient:
                  \begin{equation*}
                    \nabla_{\mathbf{x}_0}
                      \sum_{(u, v, t) \in E_t}
                      \sum_{s \in \{ -1, +1 \}}
                      q^s(u, v)
                      \log \left(
                                \frac{\kappa^s(x_{t, u}, x_{t, v})}
                                     {\sum_{v' \in V^*_t} \kappa^s(x_{t, u}, x_{t, v'}) }
                                \right)
                  \end{equation*}
                  \State \phantom{Using...} Update $\mathbf{x}_0, \mathbf{w}, \mathbf{\sigma}$ by ascending the gradient:
                  \begin{equation*}
                    \nabla_{\mathbf{x}_0, \mathbf{w}, \mathbf{\sigma}}
                      \sum_{(v, a, t) \in F_t}
                        \log \left( %
                                    \frac{\kappa_\sigma(x_{t, v}, w_a)}
                                         {\sum_{a' \in A} \kappa_\sigma(x_{t, v}, w_{a'}) }
                            \right) %
                  \end{equation*}
            \Until{convergence}
            \State $\forall (u ,v, t) \in E,\: \signfun_t(u, v, t) := \text{sign} \left( q^+(u, v) - q^-(u, v) \right)$
    \end{algorithmic}
  \end{flushleft}
\end{algorithm}

\begin{algorithm}[h!]
    \caption{Complete online learning process}
    \label{alg:complete-learning}
    \begin{flushleft}
    \algorithmicrequire \; Interaction graph $G=(V,E)$; action graph $Z=(V,A,F)$\\
    \algorithmicensure \; opinions $\mathbf{x}_0$, actions $( \mathbf{w}, \mathbf{\sigma}) $, signs $\signfun: E \rightarrow \{-1, +1\}$
    \begin{algorithmic}[1]
            \For{number of multiple restarts}
              \State Initialize $\mathbf{x}_0, \mathbf{w}, \mathbf{\sigma}$ randomly
              \For{number of epochs}
                \For{$t$ in $1, \dots, T $}
                  \State Define $\mathcal{U}^{t}$ with $(E_{< t}, \signfun_{< t})$ \Comment{Eq.~\ref{eq:defineut}}
                  \State Define $\alpha_t$ with $\mathbf{x}_t=\mathcal{U}^{t}(\mathbf{x}_0)$ \Comment{Eq.~\ref{eq:alphat}}
                  \State $\mathbf{x}_0, \mathbf{w}, \signfun_t, := \textsc{EmStep}(G_t, Z_t, \mathcal{U}^{t}(\mathbf{x}_0), \mathbf{w}, \mathbf{\sigma}, \alpha_t)$
                \EndFor
              \EndFor

              \State Keep $(\mathbf{x}_0, \mathbf{w}, \mathbf{\sigma}, \signfun)$ if $P(E,F|\mathbf{x}_0, \mathbf{w}, \mathbf{\sigma}, \signfun)$ is higher than last restart

            \EndFor
    \end{algorithmic}
  \end{flushleft}
\end{algorithm}

\end{footnotesize}

\section{Reproducibility}
\label{sec:pseudocode}
Algorithm \ref{alg:online-step} provides the pseudocode of the EM method for each time step $t$, as introduced in Section \ref{sec:algorithms}; while  Algorithm \ref{alg:complete-learning} the pseudocode of the complete learning process. Our implementation of the proposed framework, alongside all the resources needed to reproduce our experiments are available at:\\
 \url{https://github.com/corradomonti/learnable-opinion-dynamics}

\spara{Parameter settings.}
The main parameters for the evaluation are the latitudes of acceptance and contrast (\epsp, \epsn).
We fix the other parameters heuristically via grid search by optimizing the likelihood of the model.
Specifically, we use action learning rate $10^{-3}$, interaction learning rate $10^{-4}$.
In this way, we fix the the steepness of the sigmoid functions $\sigmoid_G$ used in Eq.~\ref{eq:probability-arcs} and $\sigmoid_Z$ used in Eq.~\ref{eq:feature-probability} to the values of $\rho_G = 8$ and $\rho_Z = 16$, respectively.

In the synthetic data generation, we use 30 nodes, 20 actions, 10 time steps, 3 interactions per time step per node and 15 actions per time step per node.
For the Reddit application, we fix $\mup=10^{-3}$ and $\mun=10^{-4}$. We also add to the loss function in Equation~\ref{eq:feature-likelihood} a prior on $\boldsymbol{\sigma}$ (the half-width of each $w_a$), so that it follows a $\beta(8,8)$ distribution (centered in $0.5$, with support on $[0,1]$).

\section{Linking back to $x_0$}
\label{sec:reducing-to-x0}
At each time step, the EM Algorithm updates an estimate of the same parameters: $x_0, w, \sigma$.
Thus, we need to express every $x_t$ appearing in the formulas in terms of the same parameters $x_0$, so that the gradient descent can update $x_0$.
We need therefore an efficient way to define $x_t$ in terms of $x_0$.
The opinion vector $\mathbf{x}_\tau$ is a deterministic function of $\mathbf{x}_0$ and of the signed arcs at previous time steps $(E_{< \tau}, \signfun_{< \tau})$, that we consider to be fixed.

To find a computationally efficient way to compute $\mathbf{x}_t$, we define the following $n \times n$
matrix $M$ for the signed arcs at time $t$:

\begin{equation}
  M_{u, v}(t) = \begin{cases}
      -\mu^- \cdot \#_E (u, v, t) & \text{ if } \signfun(u, v, t) = -1 \\
       \mu^+ \cdot \#_E (u, v, t) & \text{ if } \signfun(u, v, t) = +1 \\
       0 & \text{ if } (u, v, t) \notin E
  \end{cases}
\end{equation}

where $\#_E(e)$ is the multiplicity of the arc of $e$ in the multiset $E$. Then, the opinion update
(Equation~\ref{eq:opinion-update}) can be written as
\begin{footnotesize}
  \begin{multline}
    \label{eq:defineut}
        x_{t, v} + \sum_{u \in N} M_{u, v} \cdot (x_{t, u} - x_{t, v}) 
     = x_{t, v} +
        \sum_{u \in N } \Big(
           M_{u, v} \cdot x_{t, u}
        \Big) - \Big( \sum_{u \in N } M_{u, v} \Big) \cdot x_{t, v} \\
     = \Big( 1 - \sum_{u \in N} M_{u, v} \Big) x_{t, v}
         + \Big( \sum_{u \in N} M_{u, v} \cdot x_{t, u} \Big).
  \end{multline}
\end{footnotesize}

Therefore, the update equation for $\mathbf{x}_t$ can be conveniently written as a matrix operation
\begin{footnotesize}
     $\mathbf{x}_{t + 1} = %
        ( \mathbf{1} - M^\top \mathbf{1} ) \circ \mathbf{x}_t + M^\top \mathbf{x}_t$, %
\end{footnotesize}
where $\mathbf{1}$ is a vector of $|V|$ elements, $\circ$ is the Hadamard product. Let us call
$\mathcal{U}^{t}$ the repeated application of this operation, for the sequence $M(0), \dots, M(t -
1)$, applying also the clipping $\min(1, \max(0, \mathbf{x}_{t}))$ at each step. This is a
deterministic function, computed from $\signfun_{<t}, E_{<t}$, $\mu^+, \mu^-$, that gives
$\mathbf{x}_{t} = \mathcal{U}^{t}(\mathbf{x}_0) $.

\section{Notation reference}\label{sec:notation}
For readers' convenience we provide a reference table summarizing all the notation used in the paper.
\bigskip

\footnotesize

\begin{tabular}{cl}
    \toprule
		Variable & Meaning \\
	  \midrule
    $V$ & Set of actors \\
    $E$ & Interactions: $(u, v, t) \in E$ means $u$ influenced $v$ at time $t$ \\
		$G$ & Temporal graph $(V, E)$ \\
		$A$ & Set of actions \\
		$F$ & Actor-action arcs: $(v, a, t) \in F$ means $v$ performed $a$ at time $t$ \\
		$Z$ & Temporal bipartite graph $(V, A, F)$ \\
    $E_t$ & Subset of $E$ considering only arcs at time $t$ \\
    $E_{<t}$ & Subset of $E$ considering only arcs before time $t$ \\
    $G_t$ & Graph $(V, E_t)$ \\
		$F_t$ & Subset of $F$ considering only arcs at time $t$ \\
		$Z_t$ & Graph $(V, F_t)$ \\
    
    
		$\mathbf{x}_{t,v}$ & Opinion of actor $v$ at time $t$ \\
		$\mathbf{w}_a, \mathbf{\sigma}_a$ & Center and half-width of action $a$ in opinion space \\
		$\signfun$ & Sign $\{1, -1\}$ of each interaction $(u, v, t) \in E$ \\
		$\signfun_{t}$ & Restriction of $\signfun$ to $E_{t}$ \\
		$\signfun_{<t}$ & Restriction of $\signfun$ to $E_{<t}$ \\
    
    $\alpha_t$ & Probability of an interaction being positive at time $t$ \\
    $\mathcal{U}^{t}$ & Function s.t. $\mathbf{x}_{t} = \mathcal{U}^{t}(\mathbf{x}_0)$ \\
    $\epsilon^+$ & Latitude of acceptance, i.e. threshold for pos. interactions \\
    $\epsilon^-$ & Latitude of contrast, i.e. threshold for neg. interactions \\
    $\mu^+, \mu^-$ & Speed of positive and negative influence \\
    $\kappa^+, \kappa^-$ & Sigmoid function for probability of pos. and neg. interactions \\
    $\kappa_\sigma$ & Sigmoid function for probability of an actor performing an action \\
    
		\bottomrule
\end{tabular}

\end{document}